\documentclass[aps,pra,twocolumn,superscriptaddress]{revtex4-2}

\usepackage[utf8]{inputenc}
\usepackage{graphicx}
\usepackage{flafter}

\usepackage{amsmath,amssymb,amstext,amsthm,amsfonts}
\usepackage{braket}
\usepackage{bm}
\usepackage{float}
\usepackage{color}
\usepackage{subfigure}
\usepackage{blindtext}
\usepackage{marginnote}
\usepackage{ulem}

\begin{document}



\title{Tunable valley filtering in dynamically strained $\alpha$-$\mathcal{T}_3$ lattices}


\author{Alexander Filusch}
\affiliation{Institute of Physics,
University Greifswald, 17487 Greifswald, Germany }\author{Holger Fehske}
\email{fehske@physik.uni-greifswald.de}
\affiliation{Institute of Physics,
University Greifswald, 17487 Greifswald, Germany }
\affiliation{Erlangen National High Performance Computing Center, 91058 Erlangen, Germany}


\date{\today}

\begin{abstract}
	Mechanical deformations in $\alpha$-$\mathcal{T}_3$ lattices induce local pseudomagnetic fields of opposite directionality for different valleys. When this strain is equipped with a dynamical drive, it generates a complementary valley-asymmetric pseudoelectric field, which is expected to accelerate electrons.
	We propose that by combining these effects by a time-dependent nonuniform strain, tunable valley filtering devices can be engineered that extend beyond the static capabilities.
	We demonstrate this by implementing an oscillating Gaussian bump centered in a four-terminal Hall bar $\alpha$-$\mathcal{T}_3$ setup and calculating the induced pseudoelectromagnetic fields analytically. Within a recursive Floquet Green-function scheme, we determine the time-averaged transmission and valley polarization, as well as  the spatial distributions of the local density of states and current density.
	As a result of the periodic drive, we detect novel energy regimes with a highly valley-polarized transmission, depending on $\alpha$. Analyzing the spatial profiles of the time-averaged local density of states and current density we can relate these regimes to the pseudoelectromagnetic fields in the setup. 
	By means of the driving frequency, we can manipulate the valley-polarized states, which might be advantageous for future device applications. 
	
\end{abstract}

\maketitle

\section{Introduction}
\label{Intro}

In addition to charge and spin, electrons possess a valley degree of freedom in exciting two-dimensional condensed matter materials such as graphene or transition metal dichalcogenides.
For example, graphene has two inequivalent Dirac cones at the $\mathbf{K}$ and $\mathbf{K}'$ valleys of the Brillouin zone~\cite{CGPNG09}.
These valleys can be used to carry and encode information for logical operations, opening the field of 'valleytronics'~\cite{RTB07,PXYN07,SYCRRSYX16}. 
The main challenge for valleytronic devices is the precise control of valley polarization. Several proposals for valley filters and spatially separated valley-resolved currents by, e.g., nanoconstrictions~\cite{RTB07}, inversion-symmetry breaking~\cite{GSYKWCMGNLG14} or line defects~\cite{CCSFP15,JZQM11} have been made.

Graphene's outstanding ability to withstand mechanical deformations of up to 25\% due to the strong $sp^2$ bonds~\cite{LWKH08} and, most importantly, its extraordinary electromechanical coupling is particularly promising in this regard.
Because geometrical deformations modify the electronic hopping amplitude between the atoms, effective gauge fields with corresponding pseudomagnetic fields (PMFs) of more than 300 T can be generated~\cite{LBMPZGNC10}.
Remarkably, electrons residing in the $\mathbf{K}$ or $\mathbf{K}'$ valley feel opposing strain-induced PMFs since time-reversal symmetry is conserved. 
In connection, inhomogeneous PMFs due to out-of-plane deformations have attracted much attention because valley filters and beam splitters can be engineered~\cite{CFLMS14,SFVKS15,SPBJ16,MP16,TMD20,YKY22}, for an overview see Ref.~\cite{ZS19}. Experimentally, such deformations can be created and controlled by STM tips~\cite{KKZLW12}. Since the observed effects depend heavily on the energy and the degree of the deformation, there have been efforts to improve the valley-filtering efficiency by arranging multiple Gaussian bumps in superlattices~\cite{TSSSB19,GCTRP22}.
Interestingly, time-dependent strains introduce additional pseudoelectric fields (PEFs) proportional to the effective gauge field. The PEFs give rise to  valley-current generation and phonon damping~\cite{JLCKG13,OGM09} and charge pumping in mechanical resonators~\cite{WWL18}. Recently, it has been demonstrated that graphene nanodrums are viable means for valleytronic devices~\cite{OST22}. Also, crossed pseudoelectromagnetic fields have been shown to produce a charge current via a pseudo-Hall effect~\cite{SBOS20,AS21}. 
Oscillating out-of-plane strains have already been realized in nanoelectromechanical systems, where a suspended graphene membrane or ribbon is driven by an ac gate voltage with typical resonance frequencies in the MHz to GHz range~\cite{BZVFTPCM07,JRZEMS19}.

The somewhat more complicated $\alpha$-$\mathcal{T}_3$ lattice is obtained by placing an additional atom at the center of each hexagon in the honeycomb lattice with strength $\alpha$, thereby interpolating between graphene ($\alpha=0$) and the dice lattice ($\alpha=1$)~\cite{VMD98,VBDM01,RMFPM14}. 
Most notably, an additional, strictly flat band appears at zero energy going through the $\mathbf{K}$ and $\mathbf{K}'$ points, while the conduction and valence bands remain unaltered. There are several proposals for experimental realizations. The dice lattice can be manufactured by growing trilayers of cubic lattices, e.g., SrTiO$_3$/SrIrO$_3$/SrTiO$_3$, in the (111) direction~\cite{WR11}. In two dimensions, Hg$_{1-x}$Cd$_x$Te at a critical doping has been reported to map onto the $\alpha$-$\mathcal{T}_3$ lattice with an intermediate $\alpha=1/\sqrt{3}$ parameter~\cite{BUGH09,RMFPM14}. There are also several suggestions for an optical $\alpha$-$\mathcal{T}_3$ lattice that would allow a tuning of $\alpha$ by dephasing one pair of the three counter-propagating laser beams~\cite{BUGH09,RMFPM14}. Under external electromagnetic fields, the flat band and $\alpha$-dependent Berry phase have striking consequences on the  Landau level quantization~\cite{RMFPM14,FF20}, the quantum Hall effect~\cite{ICN15,BG16}, Klein tunneling~\cite{IN17,FWF20,WIZGH21,CCPFDP22} and  Weiss oscillations~\cite{FD17}. While the flat band has zero group velocity and therefore zero conductivity, it is predicted to play an important role for the transport by its nontrivial topology~\cite{WR11, NSCM11}, the coupling to propagating bands~\cite{VOVSDCD13, CXWLM19, WXHL17}, or 
interaction effects~\cite{SKOD20, WL11, NSCM11, CZZZ21}. Furthermore, electrons dressed by external laser fields or shaking in the $\alpha$-$\mathcal{T}_3$ model have attracted interest particularly in view of tuning the electronic properties~\cite{DG18,MISCN20,IZGHF22,CX22} up to the point of inducing Floquet topological phase transitions~\cite{DG19}. Accordingly, several works have exploited the valley degree of freedom, e.g., via the geometric (valley) Hall effect~\cite{XHHL17}, magnetic Fabry-P\'erot interferometry~\cite{BMCK20}, or crossed Andreev reflections~\cite{ZS22}. 

Recently, elastic deformations in $\alpha$-$\mathcal{T}_3$ structures have been shown to induce PMFs that efficiently valley filter incoming electrons by excitation to $\alpha$-dependent (pseudo) Landau levels~\cite{FBSWH21, SLDG22}. 
When the out-of-plane deformations also oscillate in time, complementary PEFs are induced that drive electrons of opposite valleys in different directions. 
So far, however, the electronic transport properties of graphene ($\alpha=0$) nanoelectromechanical systems have mostly been discussed in the adiabatic limit, i.e., the ultrafast electrons simply perceive a static deformation profile due to the slowly oscillating nanodrum~\cite{OST22}.

In this paper, we therefore consider a time-periodically oscillating Gaussian bump on an $\alpha$-$\mathcal{T}_3$ lattice within the Floquet theory and show how the PEF improves the valley-filter capabilities and affects the flat band. For this, we study the transmission of electrons in a four-terminal Hall bar setup with zigzag terminations. The advantage of such driving is the possible tuning of $\alpha$-dependent valley-polarized states by the driving frequency and the creation of (locally) flat bands for $\alpha>0$.
The paper is structured as follows. In Sec.~\ref{model_methods}, we introduce the tight-binding model with time-periodic Gaussian deformations and derive the Fourier decomposition of the time-dependent transfer amplitudes analytically. We also provide results for the DC transmission, the time-averaged local density of states  and the current density in four-terminal devices under a periodic drive in the non-equilibrium Green-function formalism. 
In Sec.~\ref{results}, we analyze a typical four-terminal transport setup and calculate all relevant quantities through a recursive Floquet Green-function algorithm built on top of the \small{\textsc{kwant}} toolbox~\cite{GWAW14}. We conclude in Sec.~\ref{Summary}.

\begin{figure}[htb]
	\centering
	\includegraphics[width=\columnwidth]{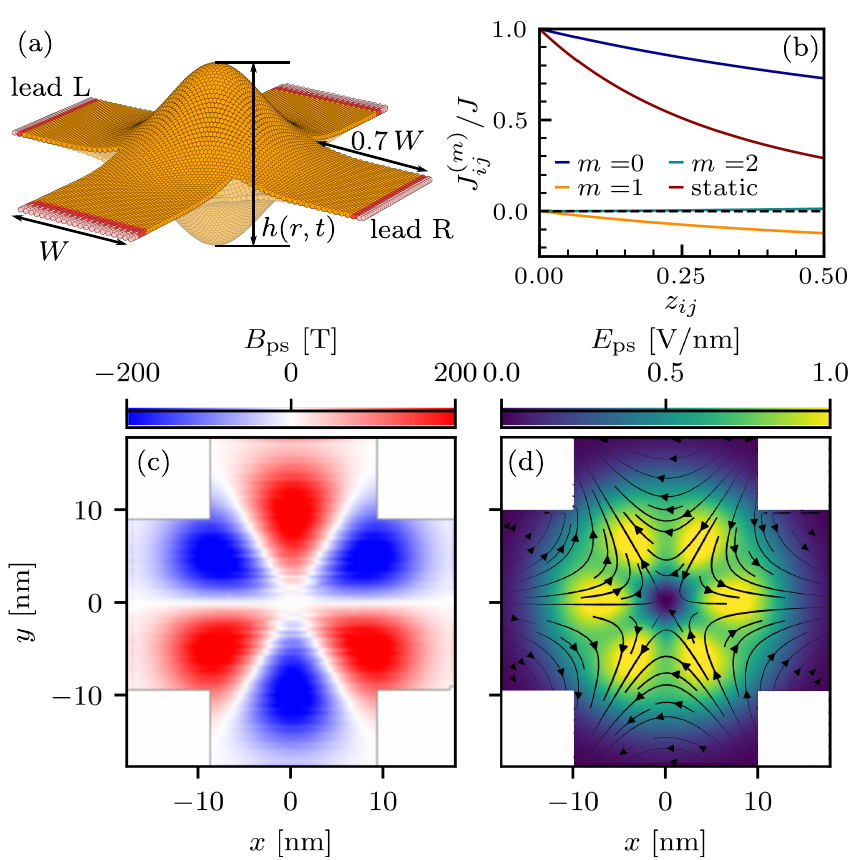}
	\caption{(a) Four-terminal setup with a dynamic Gaussian deformation~\eqref{eq:deform} in the center oscillating with frequency $\Omega$. The parameters are $h_0=8$~nm, $\sigma=10$~nm, and $W=20$~nm. Semi-infinite leads (red) attached to the Hall bar region are modeled by the pristine tight-binding Hamiltonian, i.e., $J_{ij}=J$. For the coupling between the Hall bar and the leads, we take $J_{ij}=J$ also. To create the system numerically, we use the \small{\textsc{kwant}} library~\cite{GWAW14}. (b) Comparison of the  Fourier components of the time-dependent transfer amplitudes for representative values of $z_{ij}$, and $m=0$, 1, 2. The static $J_{ij}(t=0)/J$ case is also showcased. The horizontal line denotes $J=0$. [(c) and (d)] (Zoomed-in) Pseudomagnetic and pseudoelectric field at, respectively, $t=0$ and $t=T/8$ due to the time-periodic oscillating Gaussian bump from (a) with $\Omega=0.25J$ for electrons in the $\mathbf{K}$ valley.}
	\label{fig1}
\end{figure}

\section{Model and Methods}
\label{model_methods}

To describe the electronic properties of the time-periodically strained $\alpha$-$\mathcal{T}_3$ lattice, we consider the following tight-binding Hamiltonian ($\hbar=1$)
\begin{align} 
H_\alpha(t) =& - \sum \limits_{\langle i j\rangle}J_{ij}(t) a^\dagger_i b_j^{}-  \alpha \sum\limits_{\langle i j\rangle} J_{ij}(t)  c^\dagger_i b_j^{} + \text{H.c.}\,,\label{eq:Tight-Binding}
\end{align}
where $a^{(\dagger)}$, $b^{(\dagger)}$ and $c^{(\dagger)}$ create (annihilate) an electron on Wannier sites A, B and C, respectively. 
In the $\alpha$-$\mathcal{T}_3$ lattice, an additional site C is placed at the center of each hexagon formed by the A and B sites. This site couples to the B sites via $\alpha J_{ij}(t)$, which allows for an interpolation between graphene ($\alpha=0$) and the dice lattice ($\alpha=1$). In the unstrained lattice, the nearest-neighbor hopping is $J_{ij}=J$ and the $\alpha$-$\mathcal{T}_3$ lattice features a graphene-like band structure with an additional dispersionless band at zero energy~\cite{RMFPM14}.

The out-of-plane lattice distortion $h(\mathbf{r}, t)$ alters the site positions $\mathbf{r}_i^\prime(t) = \mathbf{r}_i+ h(\mathbf{r}_i,t)\mathbf{e}_z$ [cf. Fig.\ref{fig1}(a)], thereby varying the bond length $d_{ij}(t)=|\mathbf{r}_i^\prime(t)-\mathbf{r}_j^\prime(t)|$ between nearest neighbors. The modified transfer amplitude is given by
\begin{align} 
J_{ij}(t) = J \exp\{-\beta(d_{ij}(t)/a-1)\}, \,
\label{eq:hopping}
\end{align}
where $\beta=-\partial\ln J/\partial\ln a$ is the Grüneisen parameter with $a$ denoting the (unstrained) nearest-neighbor distance.

In this work, we look upon a temporal oscillating Gaussian bump 
\begin{align}
h(r, t) = h_0 \cos(\Omega t) \exp{\left(-r^2/\sigma^2 \right)}, \label{eq:deform}
\end{align}
where $r$ is the radial distance from the center, and $h_0$ and $\sigma$ denote the bump's height and width, respectively. The Gaussian deformation is assumed to be periodic in time, $h(r,t+T)=h(r,t)$, where $T$ is the oscillation period and $\Omega=2\pi/T$ the corresponding frequency.
Note that since $h(r,t)^2$ enters  Eq. \eqref{eq:hopping},  $J_{ij}(t)$ [and thereby $H^{(\alpha)}(t)$] has a periodicity of $T/2$  (frequency of $2\Omega$) instead.

We expand Eq.~\eqref{eq:Tight-Binding} in a Fourier series, $H_\alpha(t)=\sum \limits_{m} e^{i2m\Omega t} H_\alpha^{(m)}$, with the Fourier coefficients given by $H_{\alpha}^{(m)} =\frac{2}{T}\int_0^{T/2} e^{i2m\Omega t}H_\alpha(t) \,\mathrm{d}t$. 
Rewriting Eq.~\eqref{eq:hopping} as $J_{ij}(t)/J=\exp{g(z_{ij},t)}$, where $g(z_{ij},t)=-\beta (\sqrt{1+z_{ij}\cos^2\Omega t}-1)$, we can expand Eq.~\eqref{eq:hopping} with respect to the (squared) height differences between nearest neighbors, 
\begin{align}
z_{ij}= \frac{h_0^2}{a^2}\left(e^{-r_i^2/\sigma^2}-e^{-r_j^2/\sigma^2}\right)^2,
\end{align} 
in a Taylor series around the pristine case $z_{ij}=0$. Then, using Fa\`a di Bruno's formula, we obtain
\begin{align}
 J_{ij}(t)/J = \sum_{n=0}^\infty \sum_{k=0}^n B_{n,k}(t) \frac{z_{ij}^n}{n!}, \label{eq:J(t)}
\end{align}
 where 
\begin{align}
B_{n,k}(t) = \beta^k \frac{(-1)^n}{2^n}  \cos^{2n}(\Omega t)\left[2(n-k)-1\right]!! \binom{2n-k-1}{2(n-k)} \label{eq:Bell}
\end{align}
denotes the partial Bell polynomials of the second kind for the $n$th derivative of $\exp g(z_{ij}, t)$ at $z_{ij}=0$ \cite{QSLK17}.
Due to 
\begin{align} 
\cos^{2n}(\Omega t) = \frac{1}{2^{2n}}\left\{\sum \limits_{l = 0}^{n-1} 2\binom{2n}{l}\cos[2(n-l)\Omega t]+\binom{2n}{n}\right\}, 
\end{align}
\cite{ZMGR14}, the Fourier coefficients of the time-dependent Hamiltonian
\begin{align}
H^{(m)}_\alpha =& - \sum \limits_{\langle i j\rangle}J^{(m)}_{ij} a^\dagger_i b_j^{}-  \alpha \sum\limits_{\langle i j\rangle} J^{(m)}_{ij}  c^\dagger_i b_j^{} + \text{H.c.}
\end{align}
become
\begin{align}
J^{(m)}_{ij}/J &=  \sum \limits_{n=|m|}^\infty \sum_{k=0}^n B_{n,k}(0) \binom{2n}{n-|m|}\frac{1}{2^{2n}} \frac{z_{ij}^n}{n!}\,. \label{eq:J_m}
\end{align}
\begin{figure}
	\centering
	\includegraphics[width=\columnwidth]{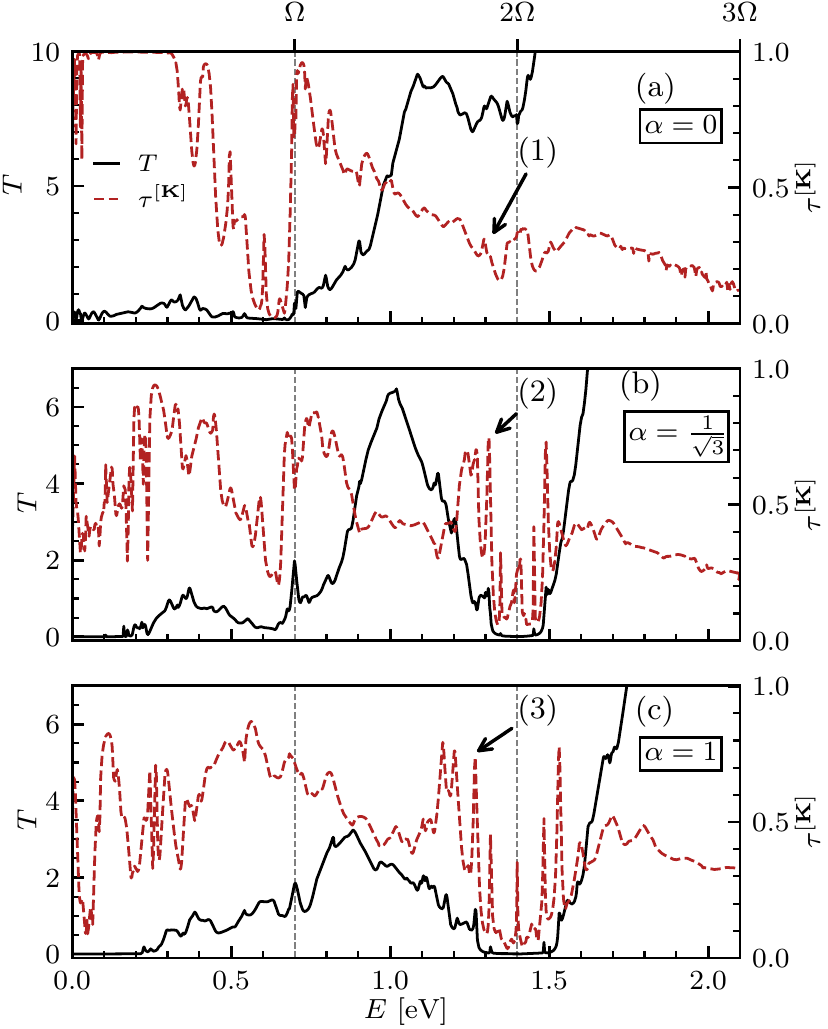}
	\caption{[(a)-(c)] Transmission and valley polarization in dependence of (quasi-) energy for an oscillating Gaussian bump in a $\alpha$-$\mathcal{T}_3$ lattice for $\alpha=0$, $1/\sqrt{3}$ and 1, respectively. Vertical dashed lines denote $E=\Omega$ and $E=2\Omega$. The resonances at energies $E\simeq 1.31$, $\simeq 1.31$ eV, and $\simeq 1.27$ eV are marked by (1), (2), and (3), respectively, where we find $\tau^{[\mathbf{K}]}= 0.4$, $\tau^{[\mathbf{K}]}=0.75$ and $\tau^{[\mathbf{K}]}=0.75$.} 
	\label{fig2}
\end{figure}
Note that the series in Eq.~\eqref{eq:J(t)} has a convergence radius of 1, which corresponds to a strained bond length $d_{ij}=\sqrt{2}a$. Since graphene is known to sustain up to 25\%~\cite{LWKH08} of elastic strain ($z_{ij}=0.5625$), any physical deformation can be correctly described.
The dependence of the different Fourier components on the driving amplitude $h_0$ is depicted in Fig.~\ref{fig1}(b), where we show Eq.~\eqref{eq:J_m} for $m=0$, 1 and 2 as a function of the expansion parameter $z_{ij}\propto h_0^2$ . We also plot the static hopping parameter ($t=0$) from Eq.~\eqref{eq:hopping}. Due to time-averaging, $J^{(0)}_{ij}/J$ is increased compared to the static Gaussian bump.  
Clearly,  $z_{ij}$ ($h_0$) directly controls the magnitude of $J^{(m)}_{ij}$. If $z_{ij}$ is constant, $J^{(m)}_{ij}$ decreases exponentially with $m$.

Dynamic elastic strain results in a time-dependent pseudoelectromagnetic vector potential $\mathbf{A}_\text{ps}= (\text{Re} A_\text{ps}, \text{Im} A_\text{ps})$~\cite{CGPNG09, PN09, SSWHB13}, where
\begin{align}
A_\mathrm{ps}(\mathbf{r_i},t) = \frac{1}{e v_\mathrm{F}} \sum \limits_{j=1}^3 J_{ij}(t) e^{-i \mathbf{K}\cdot \boldsymbol{\delta}'_{ij}}
\end{align}
and the sum is taken over the nearest neighbors. $\mathbf{K}$ denotes the corner of the Brillouin zone and $\boldsymbol{\delta}'_{ij}=\mathbf{r}_i'-\mathbf{r}'_j$ is the strained nearest-neighbor vector. The PMF is then $\mathbf{B}_\text{ps}=\boldsymbol{\nabla} \times \mathbf{A}_\text{ps}$ and the PEF is $\mathbf{E}_\text{ps} = -\partial \mathbf{A}_\text{ps}/\partial t$. The PMF and PEF induced by oscillating Gaussian strain for electrons in the $\mathbf{K}$ valley is depicted in Figs.~\ref{fig1}(c) and (d). For electrons in the $\mathbf{K}^\prime$ valley, the signs of the PMF and PEF are reversed.
\begin{figure*}[t]
	\centering
	\includegraphics[width=1\textwidth]{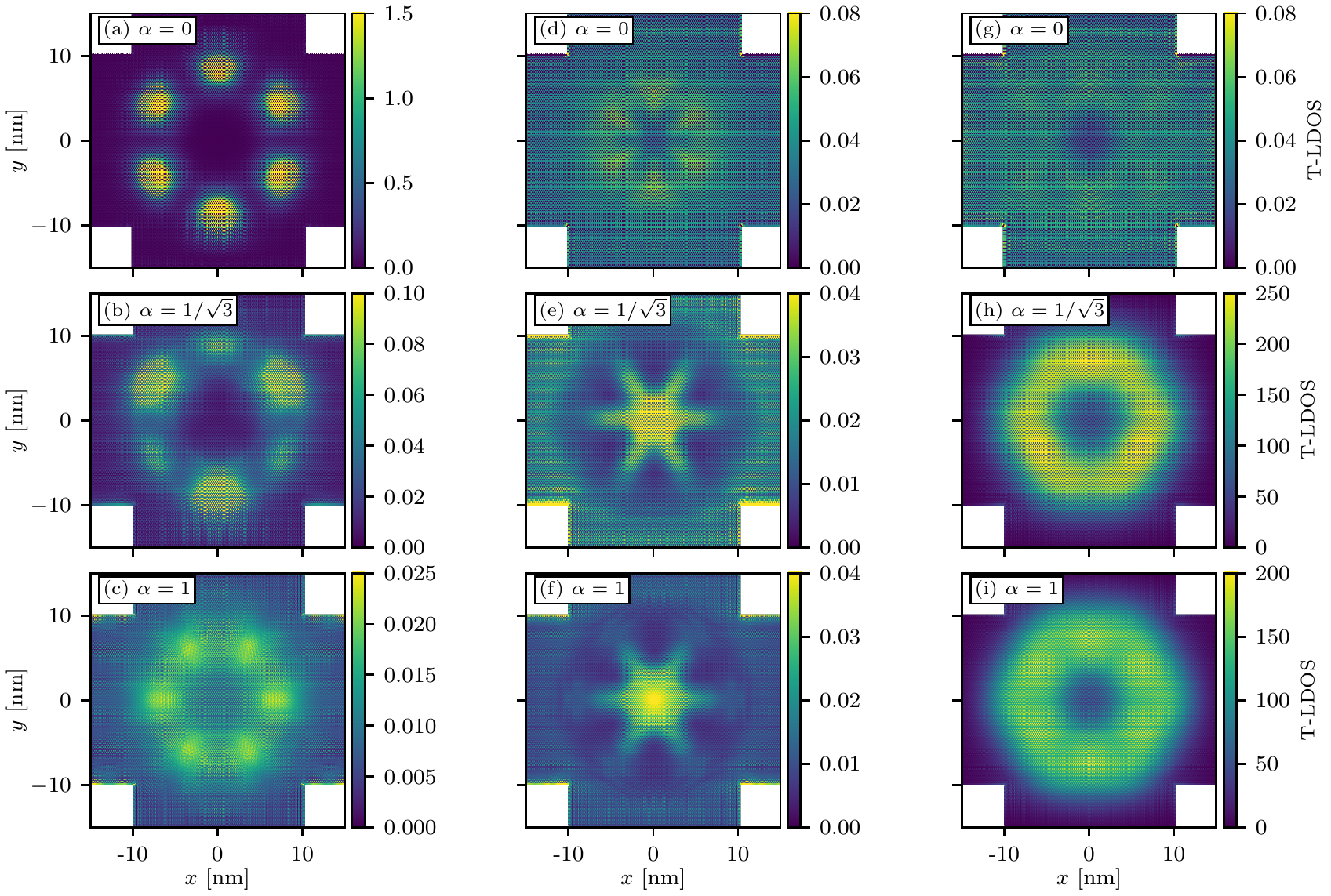}
	\caption{(Zoomed-in) Time-averaged local density of states (T-LDOS) at energies $E=\Omega$ [(a)-(c)], resonances (1-3) in [(d)-(f)] and $E=2\Omega$ [(g)-(i)]  for graphene ($\alpha=0$), the $1/\sqrt{3}$-$\mathcal{T}_3$ lattice and the dice lattice ($\alpha=1$), respectively.}
	\label{fig3}
\end{figure*}
To study the transport properties of the temporal oscillating Gaussian bump with  regard to its valley-filter capabilities, we use the four-terminal setup shown in Fig.~\ref{fig1}(a).   
The DC transmission of electrons originating from the left lead (L) to the right lead (R) is given by~\cite{KLH05, FDWWC16}
\begin{align}
T(E)  = \sum \limits_{k\in \mathbb{Z}} \text{Tr}\left[ G^r_{k 0} \Gamma^\mathrm{L}_{00} G^a_{0k} \Gamma^\mathrm{R}_{kk}\right],
\end{align}
where $G^{r(a)}_{k0}$ denotes the retarded (advanced) Green function in Floquet basis.
$\Gamma^{L(R)}_{kk} = i [\Sigma^r_{\mathrm{L(R)}} - (\Sigma^r_{\mathrm{L(R)}})^\dagger]_{kk}$ is the level-width function of the left (right) lead in Floquet representation and  $\Sigma^r_{\mathrm{L(R)}}$ denotes the respective self-energy in Floquet representation, i.e., $(\Sigma^r_\mathrm{L(R)})_{km} = \delta_{km} \Sigma^r_\mathrm{L(R)}(E+2k\Omega)$. The trace (Tr) is taken over all sites in the Hall bar region.  

The retarded Floquet Green function $G^r_{mn}$ is defined by~\cite{KOBFD11}
\begin{align}
\sum \limits_{m\in\mathbb{Z}}&\left[ (E+i\eta +2 k \Omega)\delta_{km}- H^{(k-m)}_\alpha - \left(\Sigma^r\right)_{km}\right] G^r_{mn}(E) \nonumber\\
&= \delta_{kn}, \label{eq:Floquet-GreensFunction}
\end{align}
where $(\Sigma^r)_{km}$ denotes the sum over the four lead self-energies in Floquet representation, and $i\eta$ is a small complex number guaranteeing convergence in the numerical matrix inversion. 
After truncating Eq.~\eqref{eq:Floquet-GreensFunction} at finite  $m\in [-M, M]$, we apply the recursive Green-function algorithm for Floquet systems~\cite{YZWG17, LM13} and split the four-terminal setup by a circular slicing scheme~\cite{TVE14}. Thereby, systems with up to  150 000 lattice sites with $M=6$ are accessible, which is not feasible by direct matrix inversion.

In order to study the valley-filtering, we choose zigzag boundaries for the left and right lead of the Hall bar to have well-separated valleys in momentum space and the self-energy can be projected, respectively, onto the $\mathbf{K}$ or $\mathbf{K}'$ points, i.e., $\Sigma^r_R = \Sigma^{r, [\mathbf{K}]}_R + \Sigma^{r, [\mathbf{K'}]}_R$. 
Then, 
\begin{align}
T^{[\mathbf{K}]}(E)  = \sum \limits_{k\in \mathbb{Z}} \text{Tr}\left[ G^r_{k 0} \Gamma^L_{00} G^a_{0k} \Gamma^{R, [\mathbf{K}]}_{kk}\right]
\end{align}
gives the transmission of electrons originating from the left lead into the $\mathbf{K}$ states of the right lead and we can define the valley polarization \begin{align}
\tau^{[\mathbf{K}]} = T^{[\mathbf{K}]}/T.
\end{align}
For incidents along the armchair direction, i.e., from the top or bottom lead [cf. Fig.~\ref{fig1}(a)], the Gaussian bump acts as a valley-beam splitter instead~\cite{SPBJ16}, which{would necessitate a different lead configuration~\cite{OST22,SS19}. 
In addition, the spatial distribution of the time-averaged local density of states (LDOS) on site $i$ in Floquet basis,
\begin{align}
\text{T-LDOS}(E)_i = -\frac{1}{\pi}\text{Im}\left(G^r_{00}\right)_{i,i}(E),
\end{align}
gives valuable insight into excited states inside the Gaussian bump, quite similar to the static case, where one has $\text{LDOS}_i(E) = -\frac{1}{\pi} \text{Im}\,G^r_{i,i}(E)$. The T-LDOS can be efficiently calculated by the kernel polynomial method~\cite{BWF98,WWAF06}.
\begin{figure*}[t]
	\centering
	\includegraphics[width=1\textwidth]{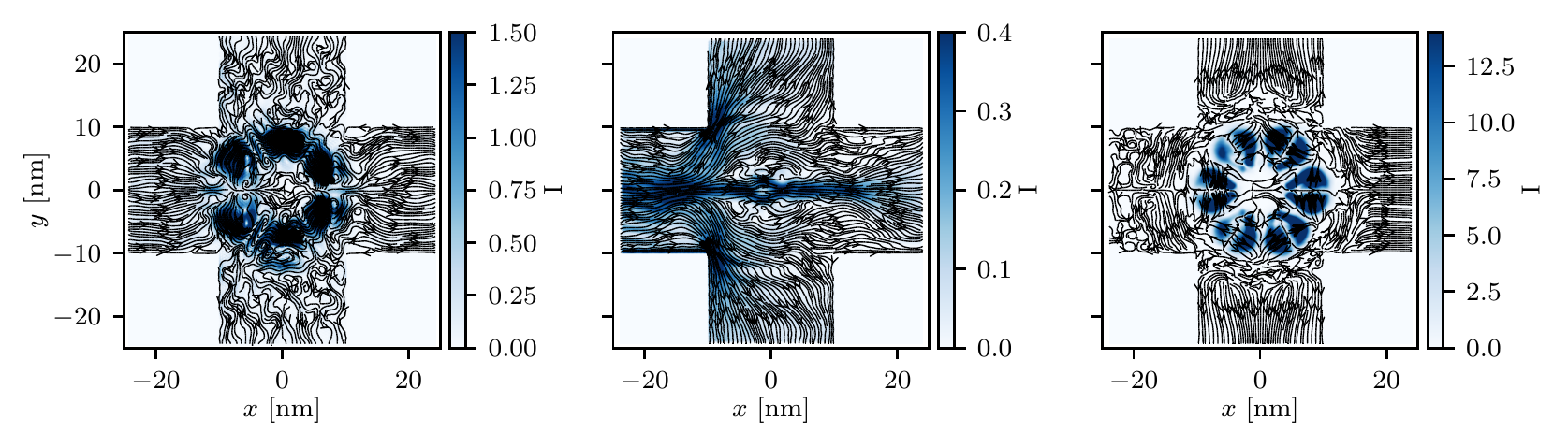}
	\caption{(Zoomed-in) Time-averaged current density profile $I$ at energies $E=\Omega$ (a), resonances (2) in (b) and $E=2\Omega$ (c)  for graphene ($\alpha=0$), the $1/\sqrt{3}$-$\mathcal{T}_3$ lattice and the dice lattice ($\alpha=1$), respectively.}
	\label{fig4}
\end{figure*}
Furthermore, the Keldysh equation connects the retarded Floquet-Green function with the lesser Floquet-Green function in Floquet representation~\cite{HJ08, WWL08},
\begin{align}
G^{<}_{mn}(E) = \sum \limits_{kl} G^r_{mk}(E) \Sigma^{<}_{kl}(E)  G^a_{ln}(E),
\end{align}
where $\Sigma^<_{kl} = i \sum \limits_p \Gamma^p_{kl}(E) f^{p}$ and $f^{p}=1/1+\exp\{(E-\mu_p)/k_\mathrm{b}T\}$ denotes the Fermi-Dirac distribution of an individual lead $p$ with chemical potential $\mu_p = E+eV_p/2$, and $V_p$ is the applied bias voltage. Then, the time-averaged and energy-resolved current density between site $i$ and site $j$ is given by~\cite{YZWG17}
\begin{align}
I_{ij}(E) = \sum \limits_{k\in \mathbb{Z}} J_{ij}^{(k)} \left[G^{<}_{0k}\right]_{ij}(E)\,.
\end{align}

\section{Results}
\label{results}

In our numerical calculations, we take the system parameters given in  Fig. \ref{fig1}, as well as the (graphene-like) values $a=0.142$ nm, $\beta=3$, and $J = 2.8$ eV for the tight-binding model~\cite{SFVKS15}, unless otherwise specified.
Let us first comment on the frequency dependency in the off-resonant case. Here, the time-reversal symmetry is intact and the time-averaged Hamiltonian $H^{(0)}_\alpha$ dominates the dynamics because the first-order corrections, $[H^{(m)}_\alpha,\,H^{(-m)}_\alpha]$, vanish in the Floquet-Magnus expansion~\cite{M54}. Concomitantly, the coupling between the Floquet sidebands is weak [cf. Fig.~\ref{fig1}(b)], necessitating frequencies smaller than the bandwidth to ensure a large overlap between neighboring sidebands.  For the parameters used, we take $J_{ij}^{(m)}$ up to $m=3$ since $z_{ij}\lesssim 0.5$  [cf. Fig.~\ref{fig1}(b)] and $M=5$ in the Floquet space to assure convergence of the truncation scheme.

Figures~\ref{fig2}(a), \ref{fig2}(b), and \ref{fig2}(c) show the transmission $T$ of electrons originating from the left lead moving to the right lead and their valley polarization $\tau^{[\mathbf{K}]}$ for graphene ($\alpha=0$), the intermediate $1/\sqrt{3}$-$\mathcal{T}_3$ lattice, and the dice lattice ($\alpha=1$), respectively. 
The main feature of the static bump is an almost complete valley polarization of the transmitted electrons stemming from the excitation to $\alpha$-dependent strain-induced Landau levels leading to "flowerlike" LDOS patterns inside the deformed region~\cite{MP16,CFLMS14, SFVKS15,FBSWH21}. One of the main caveats is the reduction in valley polarization of the output current with increasing energy since the main contribution comes from the lowest energy band~\cite{MP16}.

	For the graphene case [cf. Fig.~\ref{fig2}(a)], we notice two characteristic transport regimes, for  $E<\Omega$ and $E\simeq \Omega$.
	Starting with $E<\Omega$, we observe a plateau in the valley polarization up to $E\simeq 0.3$ eV with finite transmission because, in this energy range, only the valley-polarized zigzag edge band is occupied~\cite{TSSSB19}.
	At higher energies up to $E=\Omega$, the valley polarization decreases since higher bands are populated~\cite{MP16}. Moreover the transmission is suppressed near $E\lesssim\Omega$, indicating a gap at edge of the first Floquet zone. We found this Floquet gap irrespective of the applied frequency (below the bandwidth).
	Exactly at $E=\Omega$, we see an abrupt onset in the transmission with almost perfect valley polarization [$\tau^{[\mathbf{K}]}\simeq 0.9$] at the $m=0$ and $m=1$ zone boundary. This is quite contrary to the static case, where an increase in the energy will lower the valley polarization.

	In the intermediate case $\alpha=1/\sqrt{3}$, an additional transport channel emerges around $E\simeq 2\Omega$ [cf Fig.~\ref{fig2}(b)] related to the flat band, besides the valley-plateau at $E=\Omega$ and the static-like regime $E<\Omega$. 
	For energies below a threshold of 0.2 eV, the bump will block any current due to the large PMF inside. Above this threshold, the transmission features valley-polarized resonances with the characteristic six-fold symmetric (T-)LDOS pattern indicative of the static case. Near $E=\Omega$, we notice a similar Floquet gap with a valley-polarization plateau, albeit slightly reduced.
	Around $E\simeq 2\Omega$, we notice that the transmission vanishes (is greatly suppressed) exactly at (near) the $m=1$ zone center at $E=2\Omega$ ($E=2\Omega \pm 0.1$ eV).  Here, the flat band of the $m=1$ Floquet sideband hybridizes with the central $m=0$ band, and all states at this energy become localized [cf. discussion of Figs.~\ref{fig3}(h) and \ref{fig3}(i) below]. Since flat bands---due to their zero group velocity---do not carry any current, the transport is completely blocked. Regrettably, the numerical accuracy at the flat band heavily influences the value of the calculated polarization.  We also verified that no current is transmitted into top/bottom leads, confirming the complete blocking of any transport. The transmission gap of width $\Delta\simeq 0.18$ eV around $E=2\Omega$ is symmetric in the energy, which is a result of the particle-hole symmetric sidebands. Away from this gap we observe a small band of transmission resonances around $E\simeq 1.31$ eV [cf. (2) in Fig.~\ref{fig2}(b)] with a particular high degree of valley polarization $\tau^{[\mathbf{K}]}\simeq 0.75$. We will see that this is due to the PEF, see discussion of Figs.~\ref{fig3}(e),\ref{fig3}(f), and \ref{fig4}(b) below.
\begin{figure}
	\centering
	\includegraphics[width=1\columnwidth]{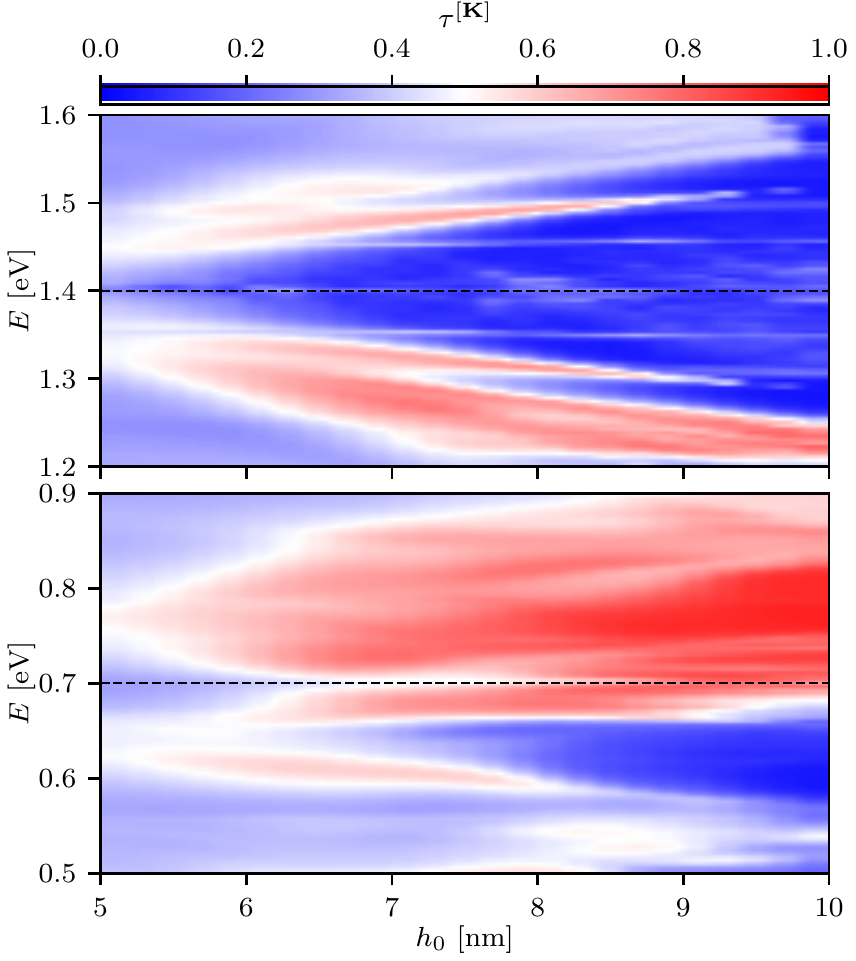}
	\caption{Contour plot of the valley polarization for the $\alpha=1/\sqrt{3}$ lattice as a function of $E$ and $h_0$ for an oscillating Gaussian bump. Horizontal dashed lines mark $E=2\Omega$ and $E=\Omega$, respectively.}
	\label{fig5}
\end{figure}
\begin{figure}
	\centering
	\includegraphics[width=\columnwidth]{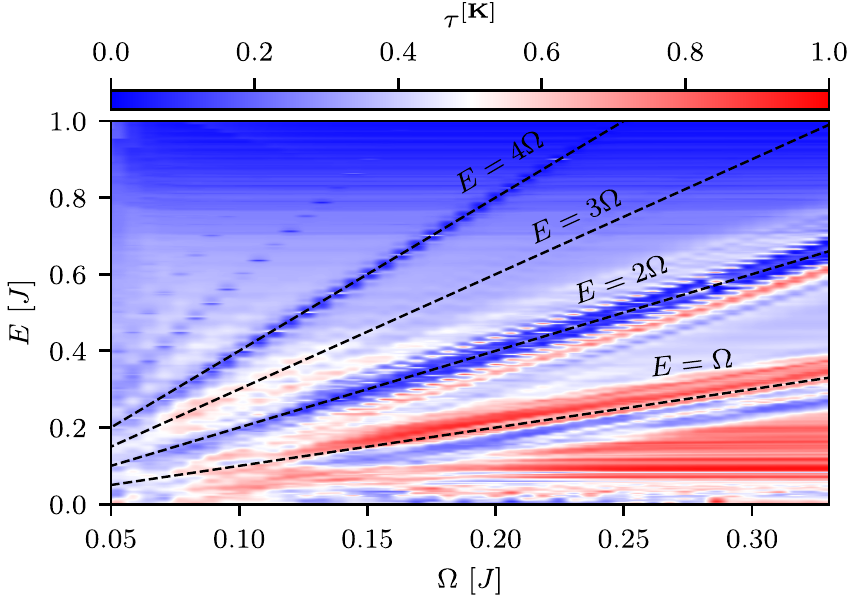}
	\caption{Contour plot of the valley polarization in dependence on (quasi-) energy and driving frequency for an oscillating Gaussian bump on the intermediate $1/\sqrt{3}$ lattice. The energy range is limited by the Van Hove singularity $E=J$ ($=2.8$~eV)  since the valley degree of freedom can only be defined up to this point. Dashed lines mark $E=m\Omega$ for $m=1$, 2, 3, and 4, respectively.}
	\label{fig6}
\end{figure}

	For the dice lattice [cf. Fig.~\ref{fig2}(c)], we can again identify the two characteristic transport regimes, $E<\Omega$ and $E\simeq 2\Omega$, of the $\alpha=1/\sqrt{3}$ model, which obviously interpolates between graphene and the dice-lattice physics. Accordingly, as necessary conditions for the different regimes, we find: When $\alpha<1$, a valley-polarization plateau is induced around $E\simeq \Omega$, while in the case of $\alpha>0$, the model features flat-band states at $E=2\Omega$ with highly valley-polarized states at the edges of the transmission gap.

	To fix the signatures of these effects and contrast them with those of the static transport channels, in Fig.~\ref{fig3} we plot the time-averaged LDOS for $E=\Omega$, resonances (1)-(3) and $E=2\Omega$  for $\alpha=0$, $1/\sqrt{3}$ and $1$, respectively.  

	At $E=\Omega$, the T-LDOS for $\alpha=0$ ($\alpha=1/\sqrt{3}$) features the sixfold  (threefold) symmetric T-LDOS pattern confined to the lobes of the PMF displayed in Fig.~\ref{fig1}(c), while the T-LDOS almost vanishes in the dice lattice case [cf. Figs.~\ref{fig3}(a)-\ref{fig3}(c)].

	For resonances (1)-(3) [cf. Fig.~\ref{fig3}(d)-\ref{fig3}(f)], we notice for $\alpha>0$ that the incoming electrons are confined to the center instead of mirroring the profile of the PMF. We attribute this to the PEF displayed in Fig.~\ref{fig1}(d), which drives the incoming electrons residing in the $\mathbf{K}$ valley to the low field region inside the bump, while electrons in $\mathbf{K}'$ valley---feeling the opposite field---are pushed away, thereby effectively polarizing the transmitted electrons [see discussion on Fig.~\ref{fig4}(b)]. This feature is characteristic for the series of resonances around (2) and (3) [and also Fig.~\ref{fig7} around (1)]. 
	The time-averaged LDOS at $E=2\Omega$ in Fig.~\ref{fig3}(h),(i) shows a highly degenerate state restricted to the Gaussian bump [cf. the color-map scale], which is in contrast to the flat band state at $E=0$, where the LDOS is spread over the whole lattice instead. In fact, the LDOS diverges at $E=2\Omega$, which indicates a flat-band like state. In the numerics, we have added a small imaginary part to the energy ($\eta=10^{-5}$) to guarantee convergence of the matrix inversion. 
	This effect occurs at $E=2m \Omega$, where $m =0, \pm 1, \pm 2, \dots $ The spectral weight increases  with increasing $h_0$ and  $1/m$. In the graphene case shown Fig.~\ref{fig3}(g), the flat band is decoupled and  the T-LDOS vanishes in accordance with the transmission calculations [cf. Fig.~\ref{fig2}(a)].

	Figures~\ref{fig4}(a), \ref{fig4}(b), and \ref{fig4}(c) provide the local (time-averaged) current densities, indicating the dynamical effects, at $E=\Omega$, resonance (2) and $E=2\Omega$, for $\alpha=0$, $1/\sqrt{3}$ and 1, respectively, where a small bias is applied between the left and right lead. The magnitude is decoded by the blue intensity, and the arrows denote the direction of the electron flow. The static-like nature of the system at $E=\Omega$ becomes particularly apparent in Fig.~\ref{fig4}(a), where the incoming stream of electrons is encircling the (distorted) lobes of the PMF, nicely representing the behavior of the T-LDOS [cf. Fig.~\ref{fig3}(a)]. The situation for resonance (2) [cf. Fig.~\ref{fig4}(b)] is much different, because, here, the PEF focuses a small part of the electron flow through the bump along the $x$-axis before exiting the scattering region through lead R. The majority of the electron flow is blocked and leaves the scattering region via the top and bottom lead, see Fig.~\ref{fig4}(b). Recalling the corresponding valley polarization in Fig.~\ref{fig2}(b), we can conclude that the transmitted stream consists primarily of $\mathbf{K}$ electrons. The density profile at $E=2\Omega$ displays a large amount of current trapped inside the bump [cf. Fig.~\ref{fig4}(c)], thereby blocking any transmission through the setup. 
	
\begin{figure}
	\centering
	\includegraphics[width=\columnwidth]{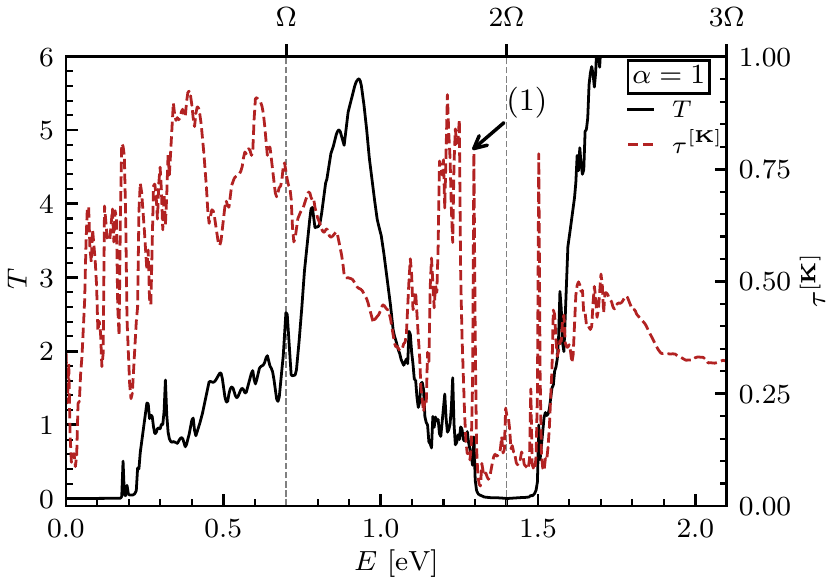}
	\caption{Transmission and valley polarization in dependence on the (quasi-) energy for an oscillating Gaussian bump on the dice lattice ($\alpha=1$) with $h_0=12$ nm, $\sigma=15$ nm, and frequency $\Omega=0.25 J$. Here, the Hall bar has $W=30$ nm. Vertical dashed lines mark $E=\Omega$ and $E=2\Omega$.}
	\label{fig7}
\end{figure}
	To assess the stability of the valley-polarized states near $E\simeq\Omega$ and $E\simeq 2\Omega$, in Fig.~\ref{fig2}(b), we provide a contour plot of the valley polarization as a function of the energy $E$  and the bump height $h_0$ in Fig.~\ref{fig5} around $E=2\Omega$ (top) and $E=\Omega$ (bottom). We again choose $\alpha=1/\sqrt{3}$ as this system best interpolates between graphene and the dice lattice. We omitted $h_0<5$ nm
	 since the amplitude, which directly controls the interband coupling between the Floquet copies [cf. Fig.~\ref{fig1}(b)],  is too weak in this case.
	In both cases, the regions of high valley polarization are robust for a wide range of amplitudes $h_0$ and we can identify  $h_0\gtrsim 6$ as the optimal regime for polarization. While the polarized resonances around $E\simeq 2\Omega$ depend almost linearly on $h_0$, the valley polarization reaches nearly unity around $E\simeq \Omega$ for large $h_0\simeq 9$~nm.

	So far, in order to avoid problems with the rapidly increasing dimension of the Floquet space, we worked at a rather large driving frequency. 
	To demonstrate that the discussed valley-polarization effects persist for smaller frequencies, in Fig.~\ref{fig6}, we show the valley polarization in the $\Omega$-$E$ plane for the $1/\sqrt{3}$ lattice. 
	Obviously, the valley-polarized regimes induced by the dynamical strain for $E=\Omega$ and $E\simeq 2\Omega$ appear down to at least $\Omega=0.1J$, albeit being slightly reduced. 
	The static-like regime $E<\Omega$ deteriorates when $\Omega < 0.2J$ through the overlap with $E=\Omega$ polarization regime. The gap $\Delta$ around $E=2\Omega$ with the valley-polarized states [cf. Fig.~\ref{fig2}(b), resonance (2)] persists until $\Delta=2\Omega$, i.e., when the resonances will be located outside the first Floquet zone. 
	 Since the discussed frequency range $34-225$ THz [$\Omega=0.05J-0.33J$] could not be expected to be realized in graphene-based nanoelectromechanical resonators~\cite{BZVFTPCM07,JRZEMS19}, we rather envision an implementation of the $\alpha$-$\mathcal{T}_3$ lattice in optical lattices, where dynamical strain also could be realized~\cite{JGL22}. The main advantage of this route is, besides having direct access to the scaling paramter $\alpha$, that the transfer amplitudes can be freely manipulated~\cite{WSOESS16,JMDLUGE14}.

	Finally, let us  comment on the system-size dependence of our results. Figure~\ref{fig7} gives  the transmission and valley polarization for a dice-lattice Hall bar  with $W=30$ nm, which corresponds to $1.5\cdot 10^5$ sites and a 50\% increase in the lead's width. We take $h_0/\sigma=0.8$ and $\sigma/W=0.5$ as in Fig.~\ref{fig1}, i.e., $h_0=12$ nm and $\sigma=15$ nm. The driving frequency remains $\Omega=0.25J$.
	We find a pronounced band of resonances around (1) $E\simeq 1.3$ eV, where the valley polarization reaches almost unity. We note that the observed effects weaken when we analyze smaller systems  (not shown). 
	In any case, it is important that the use of larger system sizes enhances the valley polarization and the corresponding resonances near $E=2\Omega$, which means that our results should be robust in the thermodynamic limit, i.e., for real-world devices. 
	This is in accord with the arguments presented in Ref.~\cite{GCTRP22}, where the authors estimated the number of pseudo-Landau level subbands by counting the iso-field orbits that enclose an integer magnetic flux. By increasing $W$, the inhomogeneous PMF is spread over a larger region, thereby increasing the number of iso-orbits (and resonant states). Although the nature of driving-induced valley-polarized states differs in our case, a similar mechanism is at work. 

\section{Summary}
\label{Summary}
In this paper, we have shown how a time-periodic modulation of strain-induced Gaussian bumps allows for engineering the transport properties in $\alpha$-$\mathcal{T}_3$ lattice Hall bars. To be more specific, we demonstrated the appearance of novel valley-polarized states and flat bands tunable by the driving frequency and the scaling parameter $\alpha$.
We discussed the role of the pseudoelectromagnetic fields in time-periodic Gaussian bumps in terms of analytical expressions for the time-dependent transfer amplitudes.
Using the recursive Floquet Green-function algorithm with a circular slicing scheme, we obtained and examined the DC transmission, valley polarization, and the spatial distribution of the time-averaged local density of states and the current density profile in a four-terminal device.
Depending on the scaling parameter $\alpha$, we identified two transport regimes with distinct valley-polarization responses caused by the dynamic strain.
For $\alpha<1$, we detected a second plateau in the valley polarization at the boundary of the central and first Floquet sideband. In the time-averaged LDOS, we found a revival of the static "flowerlike" pattern mirroring the shape of the pseudomagnetic field due to the incoming electron flow with $\mathbf{K}$ polarization encircling  the lobes of the external field. 
In the case of $\alpha>0$, we noticed a vanishing transmission at about 2$m\Omega$, which could be related to the flat bands of higher sidebands coupled to states in the zeroth sideband. The corresponding time-averaged LDOS pattern revealed a substantial spectral weight spread inside the bump, blocking any current through the device due to the zero group velocity of these flat-band states. In the vicinity of this transmission 'gap', we found highly valley-polarized resonances where the incoming ($\mathbf{K}$-polarized) electrons are focused along the zigzag orientation through the Gaussian bump by the pseudoelectric field.
We confirmed that above a threshold amplitude of the out-of-plane oscillation, the device facilitates valley filtering of the incoming electrons in both regimes, $E\simeq\Omega$ and $E\simeq 2\Omega$, with different scaling behaviors.
We enhanced the valley-filter efficiency and the resonance structure by considering larger system sizes.
While previous works on nanodrums mainly examined the adiabatic limit, we here focused on a high-frequency regime and numerically studied the influence of the external field dynamics on transport. We found pronounced effects of the PEF, e.g., $\alpha$-dependent Floquet gaps, novel valley-polarized states where the incoming electrons are focused, and flat-band states. Since Gaussian bumps in $\alpha$-$\mathcal{T}_3$ lattices have promising valley-filtering capabilities in the adiabatic and anti-adiabatic limit, we believe that observed behavior is "generic", i.e., the used high frequencies are not a restraint.  Moreover the proposed setup should be realizable in optical lattices, according to their tunability of hopping amplitudes. 


\begin{thebibliography}{74}%
\makeatletter
\providecommand \@ifxundefined [1]{%
 \@ifx{#1\undefined}
}%
\providecommand \@ifnum [1]{%
 \ifnum #1\expandafter \@firstoftwo
 \else \expandafter \@secondoftwo
 \fi
}%
\providecommand \@ifx [1]{%
 \ifx #1\expandafter \@firstoftwo
 \else \expandafter \@secondoftwo
 \fi
}%
\providecommand \natexlab [1]{#1}%
\providecommand \enquote  [1]{``#1''}%
\providecommand \bibnamefont  [1]{#1}%
\providecommand \bibfnamefont [1]{#1}%
\providecommand \citenamefont [1]{#1}%
\providecommand \href@noop [0]{\@secondoftwo}%
\providecommand \href [0]{\begingroup \@sanitize@url \@href}%
\providecommand \@href[1]{\@@startlink{#1}\@@href}%
\providecommand \@@href[1]{\endgroup#1\@@endlink}%
\providecommand \@sanitize@url [0]{\catcode `\\12\catcode `\$12\catcode
  `\&12\catcode `\#12\catcode `\^12\catcode `\_12\catcode `\%12\relax}%
\providecommand \@@startlink[1]{}%
\providecommand \@@endlink[0]{}%
\providecommand \url  [0]{\begingroup\@sanitize@url \@url }%
\providecommand \@url [1]{\endgroup\@href {#1}{\urlprefix }}%
\providecommand \urlprefix  [0]{URL }%
\providecommand \Eprint [0]{\href }%
\providecommand \doibase [0]{https://doi.org/}%
\providecommand \selectlanguage [0]{\@gobble}%
\providecommand \bibinfo  [0]{\@secondoftwo}%
\providecommand \bibfield  [0]{\@secondoftwo}%
\providecommand \translation [1]{[#1]}%
\providecommand \BibitemOpen [0]{}%
\providecommand \bibitemStop [0]{}%
\providecommand \bibitemNoStop [0]{.\EOS\space}%
\providecommand \EOS [0]{\spacefactor3000\relax}%
\providecommand \BibitemShut  [1]{\csname bibitem#1\endcsname}%
\let\auto@bib@innerbib\@empty
\bibitem [{\citenamefont {Castro~Neto}\ \emph {et~al.}(2009)\citenamefont
  {Castro~Neto}, \citenamefont {Guinea}, \citenamefont {Peres}, \citenamefont
  {Novoselov},\ and\ \citenamefont {Geim}}]{CGPNG09}%
  \BibitemOpen
  \bibfield  {author} {\bibinfo {author} {\bibfnamefont {A.~H.}\ \bibnamefont
  {Castro~Neto}}, \bibinfo {author} {\bibfnamefont {F.}~\bibnamefont {Guinea}},
  \bibinfo {author} {\bibfnamefont {N.~M.~R.}\ \bibnamefont {Peres}}, \bibinfo
  {author} {\bibfnamefont {K.~S.}\ \bibnamefont {Novoselov}},\ and\ \bibinfo
  {author} {\bibfnamefont {A.~K.}\ \bibnamefont {Geim}},\ }\href@noop {}
  {\bibfield  {journal} {\bibinfo  {journal} {Rev. Mod. Phys.}\ }\textbf
  {\bibinfo {volume} {81}},\ \bibinfo {pages} {109} (\bibinfo {year}
  {2009})}\BibitemShut {NoStop}%
\bibitem [{\citenamefont {Rycerz}\ \emph {et~al.}(2007)\citenamefont {Rycerz},
  \citenamefont {Tworzyd{\l}o},\ and\ \citenamefont {Beenakker}}]{RTB07}%
  \BibitemOpen
  \bibfield  {author} {\bibinfo {author} {\bibfnamefont {A.}~\bibnamefont
  {Rycerz}}, \bibinfo {author} {\bibfnamefont {J.}~\bibnamefont
  {Tworzyd{\l}o}},\ and\ \bibinfo {author} {\bibfnamefont {C.~W.~J.}\
  \bibnamefont {Beenakker}},\ }\href@noop {} {\bibfield  {journal} {\bibinfo
  {journal} {Europhys. Lett.}\ }\textbf {\bibinfo {volume} {79}},\ \bibinfo
  {pages} {57003} (\bibinfo {year} {2007})}\BibitemShut {NoStop}%
\bibitem [{\citenamefont {Xiao}\ \emph {et~al.}(2007)\citenamefont {Xiao},
  \citenamefont {Yao},\ and\ \citenamefont {Niu}}]{PXYN07}%
  \BibitemOpen
  \bibfield  {author} {\bibinfo {author} {\bibfnamefont {D.}~\bibnamefont
  {Xiao}}, \bibinfo {author} {\bibfnamefont {W.}~\bibnamefont {Yao}},\ and\
  \bibinfo {author} {\bibfnamefont {Q.}~\bibnamefont {Niu}},\ }\href
  {https://doi.org/10.1103/PhysRevLett.99.236809} {\bibfield  {journal}
  {\bibinfo  {journal} {Phys. Rev. Lett.}\ }\textbf {\bibinfo {volume} {99}},\
  \bibinfo {pages} {236809} (\bibinfo {year} {2007})}\BibitemShut {NoStop}%
\bibitem [{\citenamefont {Schaibley}\ \emph {et~al.}(2016)\citenamefont
  {Schaibley}, \citenamefont {Yu}, \citenamefont {Clark}, \citenamefont
  {Rivera}, \citenamefont {Ross}, \citenamefont {Seyler}, \citenamefont {Yao},\
  and\ \citenamefont {Xu}}]{SYCRRSYX16}%
  \BibitemOpen
  \bibfield  {author} {\bibinfo {author} {\bibfnamefont {J.~R.}\ \bibnamefont
  {Schaibley}}, \bibinfo {author} {\bibfnamefont {H.}~\bibnamefont {Yu}},
  \bibinfo {author} {\bibfnamefont {G.}~\bibnamefont {Clark}}, \bibinfo
  {author} {\bibfnamefont {P.}~\bibnamefont {Rivera}}, \bibinfo {author}
  {\bibfnamefont {J.~S.}\ \bibnamefont {Ross}}, \bibinfo {author}
  {\bibfnamefont {K.~L.}\ \bibnamefont {Seyler}}, \bibinfo {author}
  {\bibfnamefont {W.}~\bibnamefont {Yao}},\ and\ \bibinfo {author}
  {\bibfnamefont {X.}~\bibnamefont {Xu}},\ }\href@noop {} {\bibfield  {journal}
  {\bibinfo  {journal} {Nat. Rev. Materials}\ }\textbf {\bibinfo {volume}
  {1}},\ \bibinfo {pages} {16055} (\bibinfo {year} {2016})}\BibitemShut
  {NoStop}%
\bibitem [{\citenamefont {Gorbachev}\ \emph {et~al.}(2014)\citenamefont
  {Gorbachev}, \citenamefont {Song}, \citenamefont {Yu}, \citenamefont
  {Kretinin}, \citenamefont {Withers}, \citenamefont {Cao}, \citenamefont
  {Mishchenko}, \citenamefont {Grigorieva}, \citenamefont {Novoselov},
  \citenamefont {Levitov},\ and\ \citenamefont {Geim}}]{GSYKWCMGNLG14}%
  \BibitemOpen
  \bibfield  {author} {\bibinfo {author} {\bibfnamefont {R.~V.}\ \bibnamefont
  {Gorbachev}}, \bibinfo {author} {\bibfnamefont {J.~C.~W.}\ \bibnamefont
  {Song}}, \bibinfo {author} {\bibfnamefont {G.~L.}\ \bibnamefont {Yu}},
  \bibinfo {author} {\bibfnamefont {A.~V.}\ \bibnamefont {Kretinin}}, \bibinfo
  {author} {\bibfnamefont {F.}~\bibnamefont {Withers}}, \bibinfo {author}
  {\bibfnamefont {Y.}~\bibnamefont {Cao}}, \bibinfo {author} {\bibfnamefont
  {A.}~\bibnamefont {Mishchenko}}, \bibinfo {author} {\bibfnamefont {I.~V.}\
  \bibnamefont {Grigorieva}}, \bibinfo {author} {\bibfnamefont {K.~S.}\
  \bibnamefont {Novoselov}}, \bibinfo {author} {\bibfnamefont {L.~S.}\
  \bibnamefont {Levitov}},\ and\ \bibinfo {author} {\bibfnamefont {A.~K.}\
  \bibnamefont {Geim}},\ }\href {https://doi.org/10.1126/science.1254966}
  {\bibfield  {journal} {\bibinfo  {journal} {Science}\ }\textbf {\bibinfo
  {volume} {346}},\ \bibinfo {pages} {448} (\bibinfo {year}
  {2014})}\BibitemShut {NoStop}%
\bibitem [{\citenamefont {da~Costa}\ \emph {et~al.}(2015)\citenamefont
  {da~Costa}, \citenamefont {Chaves}, \citenamefont {Sena}, \citenamefont
  {Farias},\ and\ \citenamefont {Peeters}}]{CCSFP15}%
  \BibitemOpen
  \bibfield  {author} {\bibinfo {author} {\bibfnamefont {D.~R.}\ \bibnamefont
  {da~Costa}}, \bibinfo {author} {\bibfnamefont {A.}~\bibnamefont {Chaves}},
  \bibinfo {author} {\bibfnamefont {S.~H.~R.}\ \bibnamefont {Sena}}, \bibinfo
  {author} {\bibfnamefont {G.~A.}\ \bibnamefont {Farias}},\ and\ \bibinfo
  {author} {\bibfnamefont {F.~M.}\ \bibnamefont {Peeters}},\ }\href
  {https://doi.org/10.1103/PhysRevB.92.045417} {\bibfield  {journal} {\bibinfo
  {journal} {Phys. Rev. B}\ }\textbf {\bibinfo {volume} {92}},\ \bibinfo
  {pages} {045417} (\bibinfo {year} {2015})}\BibitemShut {NoStop}%
\bibitem [{\citenamefont {Jung}\ \emph {et~al.}(2011)\citenamefont {Jung},
  \citenamefont {Zhang}, \citenamefont {Qiao},\ and\ \citenamefont
  {MacDonald}}]{JZQM11}%
  \BibitemOpen
  \bibfield  {author} {\bibinfo {author} {\bibfnamefont {J.}~\bibnamefont
  {Jung}}, \bibinfo {author} {\bibfnamefont {F.}~\bibnamefont {Zhang}},
  \bibinfo {author} {\bibfnamefont {Z.}~\bibnamefont {Qiao}},\ and\ \bibinfo
  {author} {\bibfnamefont {A.~H.}\ \bibnamefont {MacDonald}},\ }\href
  {https://doi.org/10.1103/PhysRevB.84.075418} {\bibfield  {journal} {\bibinfo
  {journal} {Phys. Rev. B}\ }\textbf {\bibinfo {volume} {84}},\ \bibinfo
  {pages} {075418} (\bibinfo {year} {2011})}\BibitemShut {NoStop}%
\bibitem [{\citenamefont {Lee}\ \emph {et~al.}(2008)\citenamefont {Lee},
  \citenamefont {Wei}, \citenamefont {Kysar},\ and\ \citenamefont
  {Hone}}]{LWKH08}%
  \BibitemOpen
  \bibfield  {author} {\bibinfo {author} {\bibfnamefont {C.}~\bibnamefont
  {Lee}}, \bibinfo {author} {\bibfnamefont {X.}~\bibnamefont {Wei}}, \bibinfo
  {author} {\bibfnamefont {W.}~\bibnamefont {Kysar}},\ and\ \bibinfo {author}
  {\bibfnamefont {J.}~\bibnamefont {Hone}},\ }\href@noop {} {\bibfield
  {journal} {\bibinfo  {journal} {Science}\ }\textbf {\bibinfo {volume}
  {321}},\ \bibinfo {pages} {385} (\bibinfo {year} {2008})}\BibitemShut
  {NoStop}%
\bibitem [{\citenamefont {Levy}\ \emph {et~al.}(2010)\citenamefont {Levy},
  \citenamefont {Burke}, \citenamefont {Meaker}, \citenamefont {Panlasigui},
  \citenamefont {Zettl}, \citenamefont {Guinea}, \citenamefont {Neto},\ and\
  \citenamefont {Crommie}}]{LBMPZGNC10}%
  \BibitemOpen
  \bibfield  {author} {\bibinfo {author} {\bibfnamefont {N.}~\bibnamefont
  {Levy}}, \bibinfo {author} {\bibfnamefont {S.}~\bibnamefont {Burke}},
  \bibinfo {author} {\bibfnamefont {K.~L.}\ \bibnamefont {Meaker}}, \bibinfo
  {author} {\bibfnamefont {N.}~\bibnamefont {Panlasigui}}, \bibinfo {author}
  {\bibfnamefont {A.}~\bibnamefont {Zettl}}, \bibinfo {author} {\bibfnamefont
  {F.}~\bibnamefont {Guinea}}, \bibinfo {author} {\bibfnamefont
  {A.}~\bibnamefont {Neto}},\ and\ \bibinfo {author} {\bibfnamefont
  {M.}~\bibnamefont {Crommie}},\ }\href@noop {} {\bibfield  {journal} {\bibinfo
   {journal} {Science}\ }\textbf {\bibinfo {volume} {329}},\ \bibinfo {pages}
  {544} (\bibinfo {year} {2010})}\BibitemShut {NoStop}%
\bibitem [{\citenamefont {Carrillo-Bastos}\ \emph {et~al.}(2014)\citenamefont
  {Carrillo-Bastos}, \citenamefont {Faria}, \citenamefont {Latg\'e},
  \citenamefont {Mireles},\ and\ \citenamefont {Sandler}}]{CFLMS14}%
  \BibitemOpen
  \bibfield  {author} {\bibinfo {author} {\bibfnamefont {R.}~\bibnamefont
  {Carrillo-Bastos}}, \bibinfo {author} {\bibfnamefont {D.}~\bibnamefont
  {Faria}}, \bibinfo {author} {\bibfnamefont {A.}~\bibnamefont {Latg\'e}},
  \bibinfo {author} {\bibfnamefont {F.}~\bibnamefont {Mireles}},\ and\ \bibinfo
  {author} {\bibfnamefont {N.}~\bibnamefont {Sandler}},\ }\href
  {https://doi.org/10.1103/PhysRevB.90.041411} {\bibfield  {journal} {\bibinfo
  {journal} {Phys. Rev. B}\ }\textbf {\bibinfo {volume} {90}},\ \bibinfo
  {pages} {041411(R)} (\bibinfo {year} {2014})}\BibitemShut {NoStop}%
\bibitem [{\citenamefont {Schneider}\ \emph {et~al.}(2015)\citenamefont
  {Schneider}, \citenamefont {Faria}, \citenamefont {Viola~Kusminskiy},\ and\
  \citenamefont {Sandler}}]{SFVKS15}%
  \BibitemOpen
  \bibfield  {author} {\bibinfo {author} {\bibfnamefont {M.}~\bibnamefont
  {Schneider}}, \bibinfo {author} {\bibfnamefont {D.}~\bibnamefont {Faria}},
  \bibinfo {author} {\bibfnamefont {S.}~\bibnamefont {Viola~Kusminskiy}},\ and\
  \bibinfo {author} {\bibfnamefont {N.}~\bibnamefont {Sandler}},\ }\href
  {https://doi.org/10.1103/PhysRevB.91.161407} {\bibfield  {journal} {\bibinfo
  {journal} {Phys. Rev. B}\ }\textbf {\bibinfo {volume} {91}},\ \bibinfo
  {pages} {161407\mbox{(R)}} (\bibinfo {year} {2015})}\BibitemShut {NoStop}%
\bibitem [{\citenamefont {Settnes}\ \emph {et~al.}(2016)\citenamefont
  {Settnes}, \citenamefont {Power}, \citenamefont {Brandbyge},\ and\
  \citenamefont {Jauho}}]{SPBJ16}%
  \BibitemOpen
  \bibfield  {author} {\bibinfo {author} {\bibfnamefont {M.}~\bibnamefont
  {Settnes}}, \bibinfo {author} {\bibfnamefont {S.~R.}\ \bibnamefont {Power}},
  \bibinfo {author} {\bibfnamefont {M.}~\bibnamefont {Brandbyge}},\ and\
  \bibinfo {author} {\bibfnamefont {A.-P.}\ \bibnamefont {Jauho}},\ }\href
  {https://doi.org/10.1103/PhysRevLett.117.276801} {\bibfield  {journal}
  {\bibinfo  {journal} {Phys. Rev. Lett.}\ }\textbf {\bibinfo {volume} {117}},\
  \bibinfo {pages} {276801} (\bibinfo {year} {2016})}\BibitemShut {NoStop}%
\bibitem [{\citenamefont {Milovanovi^^c4^^87}\ and\ \citenamefont
  {Peeters}(2016)}]{MP16}%
  \BibitemOpen
  \bibfield  {author} {\bibinfo {author} {\bibfnamefont {S.~P.}\ \bibnamefont
  {Milovanovi^^c4^^87}}\ and\ \bibinfo {author} {\bibfnamefont {F.~M.}\
  \bibnamefont {Peeters}},\ }\href {https://doi.org/10.1063/1.4967977}
  {\bibfield  {journal} {\bibinfo  {journal} {Applied Physics Letters}\
  }\textbf {\bibinfo {volume} {109}},\ \bibinfo {pages} {203108} (\bibinfo
  {year} {2016})}\BibitemShut {NoStop}%
\bibitem [{\citenamefont {Tran}\ \emph {et~al.}(2020)\citenamefont {Tran},
  \citenamefont {Saint-Martin},\ and\ \citenamefont {Dollfus}}]{TMD20}%
  \BibitemOpen
  \bibfield  {author} {\bibinfo {author} {\bibfnamefont {V.-T.}\ \bibnamefont
  {Tran}}, \bibinfo {author} {\bibfnamefont {J.}~\bibnamefont {Saint-Martin}},\
  and\ \bibinfo {author} {\bibfnamefont {P.}~\bibnamefont {Dollfus}},\ }\href
  {https://doi.org/10.1103/PhysRevB.102.075425} {\bibfield  {journal} {\bibinfo
   {journal} {Phys. Rev. B}\ }\textbf {\bibinfo {volume} {102}},\ \bibinfo
  {pages} {075425} (\bibinfo {year} {2020})}\BibitemShut {NoStop}%
\bibitem [{\citenamefont {Yu}\ \emph {et~al.}(2022)\citenamefont {Yu},
  \citenamefont {Kutana},\ and\ \citenamefont {Yakobson}}]{YKY22}%
  \BibitemOpen
  \bibfield  {author} {\bibinfo {author} {\bibfnamefont {H.}~\bibnamefont
  {Yu}}, \bibinfo {author} {\bibfnamefont {A.}~\bibnamefont {Kutana}},\ and\
  \bibinfo {author} {\bibfnamefont {B.~I.}\ \bibnamefont {Yakobson}},\ }\href
  {https://doi.org/10.1021/acs.nanolett.2c00103} {\bibfield  {journal}
  {\bibinfo  {journal} {Nano Letters}\ }\textbf {\bibinfo {volume} {22}},\
  \bibinfo {pages} {2934} (\bibinfo {year} {2022})}\BibitemShut {NoStop}%
\bibitem [{\citenamefont {Zhai}\ and\ \citenamefont {Sandler}(2019)}]{ZS19}%
  \BibitemOpen
  \bibfield  {author} {\bibinfo {author} {\bibfnamefont {D.}~\bibnamefont
  {Zhai}}\ and\ \bibinfo {author} {\bibfnamefont {N.}~\bibnamefont {Sandler}},\
  }\href {https://doi.org/10.1142/S0217984919300011} {\bibfield  {journal}
  {\bibinfo  {journal} {Modern Physics Letters B}\ }\textbf {\bibinfo {volume}
  {33}},\ \bibinfo {pages} {1930001} (\bibinfo {year} {2019})}\BibitemShut
  {NoStop}%
\bibitem [{\citenamefont {Klimov}\ \emph {et~al.}(2012)\citenamefont {Klimov},
  \citenamefont {Jung}, \citenamefont {Zhu}, \citenamefont {Li}, \citenamefont
  {Wright}, \citenamefont {Solares}, \citenamefont {Newell}, \citenamefont
  {Zhitenev},\ and\ \citenamefont {Stroscio}}]{KKZLW12}%
  \BibitemOpen
  \bibfield  {author} {\bibinfo {author} {\bibfnamefont {N.~N.}\ \bibnamefont
  {Klimov}}, \bibinfo {author} {\bibfnamefont {S.}~\bibnamefont {Jung}},
  \bibinfo {author} {\bibfnamefont {S.}~\bibnamefont {Zhu}}, \bibinfo {author}
  {\bibfnamefont {T.}~\bibnamefont {Li}}, \bibinfo {author} {\bibfnamefont
  {C.~A.}\ \bibnamefont {Wright}}, \bibinfo {author} {\bibfnamefont {S.~D.}\
  \bibnamefont {Solares}}, \bibinfo {author} {\bibfnamefont {D.~B.}\
  \bibnamefont {Newell}}, \bibinfo {author} {\bibfnamefont {N.~B.}\
  \bibnamefont {Zhitenev}},\ and\ \bibinfo {author} {\bibfnamefont {J.~A.}\
  \bibnamefont {Stroscio}},\ }\href@noop {} {\bibfield  {journal} {\bibinfo
  {journal} {Science}\ }\textbf {\bibinfo {volume} {336}},\ \bibinfo {pages}
  {1557} (\bibinfo {year} {2012})}\BibitemShut {NoStop}%
\bibitem [{\citenamefont {Torres}\ \emph {et~al.}(2019)\citenamefont {Torres},
  \citenamefont {Silva}, \citenamefont {de~Souza}, \citenamefont {Silva},\ and\
  \citenamefont {Bahamon}}]{TSSSB19}%
  \BibitemOpen
  \bibfield  {author} {\bibinfo {author} {\bibfnamefont {V.}~\bibnamefont
  {Torres}}, \bibinfo {author} {\bibfnamefont {P.}~\bibnamefont {Silva}},
  \bibinfo {author} {\bibfnamefont {E.~A.~T.}\ \bibnamefont {de~Souza}},
  \bibinfo {author} {\bibfnamefont {L.~A.}\ \bibnamefont {Silva}},\ and\
  \bibinfo {author} {\bibfnamefont {D.~A.}\ \bibnamefont {Bahamon}},\ }\href
  {https://doi.org/10.1103/PhysRevB.100.205411} {\bibfield  {journal} {\bibinfo
   {journal} {Phys. Rev. B}\ }\textbf {\bibinfo {volume} {100}},\ \bibinfo
  {pages} {205411} (\bibinfo {year} {2019})}\BibitemShut {NoStop}%
\bibitem [{\citenamefont {Giambastiani}\ \emph {et~al.}(2022)\citenamefont
  {Giambastiani}, \citenamefont {Colangelo}, \citenamefont {Tredicucci},
  \citenamefont {Roddaro},\ and\ \citenamefont {Pitanti}}]{GCTRP22}%
  \BibitemOpen
  \bibfield  {author} {\bibinfo {author} {\bibfnamefont {D.}~\bibnamefont
  {Giambastiani}}, \bibinfo {author} {\bibfnamefont {F.}~\bibnamefont
  {Colangelo}}, \bibinfo {author} {\bibfnamefont {A.}~\bibnamefont
  {Tredicucci}}, \bibinfo {author} {\bibfnamefont {S.}~\bibnamefont
  {Roddaro}},\ and\ \bibinfo {author} {\bibfnamefont {A.}~\bibnamefont
  {Pitanti}},\ }\href {https://doi.org/10.1063/5.0080098} {\bibfield  {journal}
  {\bibinfo  {journal} {Journal of Applied Physics}\ }\textbf {\bibinfo
  {volume} {131}},\ \bibinfo {pages} {085103} (\bibinfo {year}
  {2022})}\BibitemShut {NoStop}%
\bibitem [{\citenamefont {Jiang}\ \emph {et~al.}(2013)\citenamefont {Jiang},
  \citenamefont {Low}, \citenamefont {Chang}, \citenamefont {Katsnelson},\ and\
  \citenamefont {Guinea}}]{JLCKG13}%
  \BibitemOpen
  \bibfield  {author} {\bibinfo {author} {\bibfnamefont {Y.}~\bibnamefont
  {Jiang}}, \bibinfo {author} {\bibfnamefont {T.}~\bibnamefont {Low}}, \bibinfo
  {author} {\bibfnamefont {K.}~\bibnamefont {Chang}}, \bibinfo {author}
  {\bibfnamefont {M.~I.}\ \bibnamefont {Katsnelson}},\ and\ \bibinfo {author}
  {\bibfnamefont {F.}~\bibnamefont {Guinea}},\ }\href
  {https://doi.org/10.1103/PhysRevLett.110.046601} {\bibfield  {journal}
  {\bibinfo  {journal} {Phys. Rev. Lett.}\ }\textbf {\bibinfo {volume} {110}},\
  \bibinfo {pages} {046601} (\bibinfo {year} {2013})}\BibitemShut {NoStop}%
\bibitem [{\citenamefont {von Oppen}\ \emph {et~al.}(2009)\citenamefont {von
  Oppen}, \citenamefont {Guinea},\ and\ \citenamefont {Mariani}}]{OGM09}%
  \BibitemOpen
  \bibfield  {author} {\bibinfo {author} {\bibfnamefont {F.}~\bibnamefont {von
  Oppen}}, \bibinfo {author} {\bibfnamefont {F.}~\bibnamefont {Guinea}},\ and\
  \bibinfo {author} {\bibfnamefont {E.}~\bibnamefont {Mariani}},\ }\href
  {https://doi.org/10.1103/PhysRevB.80.075420} {\bibfield  {journal} {\bibinfo
  {journal} {Phys. Rev. B}\ }\textbf {\bibinfo {volume} {80}},\ \bibinfo
  {pages} {075420} (\bibinfo {year} {2009})}\BibitemShut {NoStop}%
\bibitem [{\citenamefont {Wang}\ \emph {et~al.}(2018)\citenamefont {Wang},
  \citenamefont {Wang},\ and\ \citenamefont {Liu}}]{WWL18}%
  \BibitemOpen
  \bibfield  {author} {\bibinfo {author} {\bibfnamefont {M.~J.}\ \bibnamefont
  {Wang}}, \bibinfo {author} {\bibfnamefont {J.}~\bibnamefont {Wang}},\ and\
  \bibinfo {author} {\bibfnamefont {J.~F.}\ \bibnamefont {Liu}},\ }\href
  {https://doi.org/10.1209/0295-5075/121/47002} {\bibfield  {journal} {\bibinfo
   {journal} {{EPL} (Europhysics Letters)}\ }\textbf {\bibinfo {volume}
  {121}},\ \bibinfo {pages} {47002} (\bibinfo {year} {2018})}\BibitemShut
  {NoStop}%
\bibitem [{\citenamefont {Ortiz}\ \emph {et~al.}(2022)\citenamefont {Ortiz},
  \citenamefont {Szpak},\ and\ \citenamefont {Stegmann}}]{OST22}%
  \BibitemOpen
  \bibfield  {author} {\bibinfo {author} {\bibfnamefont {W.}~\bibnamefont
  {Ortiz}}, \bibinfo {author} {\bibfnamefont {N.}~\bibnamefont {Szpak}},\ and\
  \bibinfo {author} {\bibfnamefont {T.}~\bibnamefont {Stegmann}},\ }\href
  {https://doi.org/10.1103/PhysRevB.106.035416} {\bibfield  {journal} {\bibinfo
   {journal} {Phys. Rev. B}\ }\textbf {\bibinfo {volume} {106}},\ \bibinfo
  {pages} {035416} (\bibinfo {year} {2022})}\BibitemShut {NoStop}%
\bibitem [{\citenamefont {Sela}\ \emph {et~al.}(2020)\citenamefont {Sela},
  \citenamefont {Bloch}, \citenamefont {von Oppen},\ and\ \citenamefont
  {Shalom}}]{SBOS20}%
  \BibitemOpen
  \bibfield  {author} {\bibinfo {author} {\bibfnamefont {E.}~\bibnamefont
  {Sela}}, \bibinfo {author} {\bibfnamefont {Y.}~\bibnamefont {Bloch}},
  \bibinfo {author} {\bibfnamefont {F.}~\bibnamefont {von Oppen}},\ and\
  \bibinfo {author} {\bibfnamefont {M.~B.}\ \bibnamefont {Shalom}},\ }\href
  {https://doi.org/10.1103/PhysRevLett.124.026602} {\bibfield  {journal}
  {\bibinfo  {journal} {Phys. Rev. Lett.}\ }\textbf {\bibinfo {volume} {124}},\
  \bibinfo {pages} {026602} (\bibinfo {year} {2020})}\BibitemShut {NoStop}%
\bibitem [{\citenamefont {Amasay}\ and\ \citenamefont {Sela}(2021)}]{AS21}%
  \BibitemOpen
  \bibfield  {author} {\bibinfo {author} {\bibfnamefont {J.}~\bibnamefont
  {Amasay}}\ and\ \bibinfo {author} {\bibfnamefont {E.}~\bibnamefont {Sela}},\
  }\href {https://doi.org/10.1103/PhysRevB.104.125428} {\bibfield  {journal}
  {\bibinfo  {journal} {Phys. Rev. B}\ }\textbf {\bibinfo {volume} {104}},\
  \bibinfo {pages} {125428} (\bibinfo {year} {2021})}\BibitemShut {NoStop}%
\bibitem [{\citenamefont {Bunch}\ \emph {et~al.}(2007)\citenamefont {Bunch},
  \citenamefont {van~der Zande}, \citenamefont {Verbridge}, \citenamefont
  {Frank}, \citenamefont {Tanenbaum}, \citenamefont {Parpia}, \citenamefont
  {Craighead},\ and\ \citenamefont {McEuen}}]{BZVFTPCM07}%
  \BibitemOpen
  \bibfield  {author} {\bibinfo {author} {\bibfnamefont {J.~S.}\ \bibnamefont
  {Bunch}}, \bibinfo {author} {\bibfnamefont {A.~M.}\ \bibnamefont {van~der
  Zande}}, \bibinfo {author} {\bibfnamefont {S.~S.}\ \bibnamefont {Verbridge}},
  \bibinfo {author} {\bibfnamefont {I.~W.}\ \bibnamefont {Frank}}, \bibinfo
  {author} {\bibfnamefont {D.~M.}\ \bibnamefont {Tanenbaum}}, \bibinfo {author}
  {\bibfnamefont {J.~M.}\ \bibnamefont {Parpia}}, \bibinfo {author}
  {\bibfnamefont {H.~G.}\ \bibnamefont {Craighead}},\ and\ \bibinfo {author}
  {\bibfnamefont {P.~L.}\ \bibnamefont {McEuen}},\ }\href
  {https://doi.org/10.1126/science.1136836} {\bibfield  {journal} {\bibinfo
  {journal} {Science}\ }\textbf {\bibinfo {volume} {315}},\ \bibinfo {pages}
  {490} (\bibinfo {year} {2007})}\BibitemShut {NoStop}%
\bibitem [{\citenamefont {Vidal}\ \emph {et~al.}(1998)\citenamefont {Vidal},
  \citenamefont {Mosseri},\ and\ \citenamefont {Dou\ifmmode~\mbox{\c{c}}\else
  \c{c}\fi{}ot}}]{VMD98}%
  \BibitemOpen
  \bibfield  {author} {\bibinfo {author} {\bibfnamefont {J.}~\bibnamefont
  {Vidal}}, \bibinfo {author} {\bibfnamefont {R.}~\bibnamefont {Mosseri}},\
  and\ \bibinfo {author} {\bibfnamefont {B.}~\bibnamefont
  {Dou\ifmmode~\mbox{\c{c}}\else \c{c}\fi{}ot}},\ }\href@noop {} {\bibfield
  {journal} {\bibinfo  {journal} {Phys. Rev. Lett.}\ }\textbf {\bibinfo
  {volume} {81}},\ \bibinfo {pages} {5888} (\bibinfo {year}
  {1998})}\BibitemShut {NoStop}%
\bibitem [{\citenamefont {Vidal}\ \emph {et~al.}(2001)\citenamefont {Vidal},
  \citenamefont {Butaud}, \citenamefont {Dou\ifmmode~\mbox{\c{c}}\else
  \c{c}\fi{}ot},\ and\ \citenamefont {Mosseri}}]{VBDM01}%
  \BibitemOpen
  \bibfield  {author} {\bibinfo {author} {\bibfnamefont {J.}~\bibnamefont
  {Vidal}}, \bibinfo {author} {\bibfnamefont {P.}~\bibnamefont {Butaud}},
  \bibinfo {author} {\bibfnamefont {B.}~\bibnamefont
  {Dou\ifmmode~\mbox{\c{c}}\else \c{c}\fi{}ot}},\ and\ \bibinfo {author}
  {\bibfnamefont {R.}~\bibnamefont {Mosseri}},\ }\href
  {https://doi.org/10.1103/PhysRevB.64.155306} {\bibfield  {journal} {\bibinfo
  {journal} {Phys. Rev. B}\ }\textbf {\bibinfo {volume} {64}},\ \bibinfo
  {pages} {155306} (\bibinfo {year} {2001})}\BibitemShut {NoStop}%
  \bibitem [{\citenamefont {Jung}\ \emph {et~al.}(2019)\citenamefont {Jung},
  \citenamefont {Rickhaus}, \citenamefont {Zihlmann}, \citenamefont {Eichler},
  \citenamefont {Makk},\ and\ \citenamefont {Sch^^c3^^b6nenberger}}]{JRZEMS19}%
  \BibitemOpen
  \bibfield  {author} {\bibinfo {author} {\bibfnamefont {M.}~\bibnamefont
  {Jung}}, \bibinfo {author} {\bibfnamefont {P.}~\bibnamefont {Rickhaus}},
  \bibinfo {author} {\bibfnamefont {S.}~\bibnamefont {Zihlmann}}, \bibinfo
  {author} {\bibfnamefont {A.}~\bibnamefont {Eichler}}, \bibinfo {author}
  {\bibfnamefont {P.}~\bibnamefont {Makk}},\ and\ \bibinfo {author}
  {\bibfnamefont {C.}~\bibnamefont {Sch^^c3^^b6nenberger}},\ }\href
  {https://doi.org/10.1039/C8NR09963D} {\bibfield  {journal} {\bibinfo
  {journal} {Nanoscale}\ }\textbf {\bibinfo {volume} {11}},\ \bibinfo {pages}
  {4355} (\bibinfo {year} {2019})}\BibitemShut {NoStop}%
\bibitem [{\citenamefont {Raoux}\ \emph {et~al.}(2014)\citenamefont {Raoux},
  \citenamefont {Morigi}, \citenamefont {Fuchs}, \citenamefont {Pi\'echon},\
  and\ \citenamefont {Montambaux}}]{RMFPM14}%
  \BibitemOpen
  \bibfield  {author} {\bibinfo {author} {\bibfnamefont {A.}~\bibnamefont
  {Raoux}}, \bibinfo {author} {\bibfnamefont {M.}~\bibnamefont {Morigi}},
  \bibinfo {author} {\bibfnamefont {J.-N.}\ \bibnamefont {Fuchs}}, \bibinfo
  {author} {\bibfnamefont {F.}~\bibnamefont {Pi\'echon}},\ and\ \bibinfo
  {author} {\bibfnamefont {G.}~\bibnamefont {Montambaux}},\ }\href
  {https://doi.org/10.1103/PhysRevLett.112.026402} {\bibfield  {journal}
  {\bibinfo  {journal} {Phys. Rev. Lett.}\ }\textbf {\bibinfo {volume} {112}},\
  \bibinfo {pages} {026402} (\bibinfo {year} {2014})}\BibitemShut {NoStop}%
\bibitem [{\citenamefont {Wang}\ and\ \citenamefont {Ran}(2011)}]{WR11}%
  \BibitemOpen
  \bibfield  {author} {\bibinfo {author} {\bibfnamefont {F.}~\bibnamefont
  {Wang}}\ and\ \bibinfo {author} {\bibfnamefont {Y.}~\bibnamefont {Ran}},\
  }\href {https://doi.org/10.1103/PhysRevB.84.241103} {\bibfield  {journal}
  {\bibinfo  {journal} {Phys. Rev. B}\ }\textbf {\bibinfo {volume} {84}},\
  \bibinfo {pages} {241103} (\bibinfo {year} {2011})}\BibitemShut {NoStop}%
\bibitem [{\citenamefont {Bercioux}\ \emph {et~al.}(2009)\citenamefont
  {Bercioux}, \citenamefont {Urban}, \citenamefont {Grabert},\ and\
  \citenamefont {H\"ausler}}]{BUGH09}%
  \BibitemOpen
  \bibfield  {author} {\bibinfo {author} {\bibfnamefont {D.}~\bibnamefont
  {Bercioux}}, \bibinfo {author} {\bibfnamefont {D.~F.}\ \bibnamefont {Urban}},
  \bibinfo {author} {\bibfnamefont {H.}~\bibnamefont {Grabert}},\ and\ \bibinfo
  {author} {\bibfnamefont {W.}~\bibnamefont {H\"ausler}},\ }\href
  {https://doi.org/10.1103/PhysRevA.80.063603} {\bibfield  {journal} {\bibinfo
  {journal} {Phys. Rev. A}\ }\textbf {\bibinfo {volume} {80}},\ \bibinfo
  {pages} {063603} (\bibinfo {year} {2009})}\BibitemShut {NoStop}%
\bibitem [{\citenamefont {Filusch}\ and\ \citenamefont {Fehske}(2020)}]{FF20}%
  \BibitemOpen
  \bibfield  {author} {\bibinfo {author} {\bibfnamefont {A.}~\bibnamefont
  {Filusch}}\ and\ \bibinfo {author} {\bibfnamefont {H.}~\bibnamefont
  {Fehske}},\ }\href@noop {} {\bibfield  {journal} {\bibinfo  {journal} {Eur.
  Phys. J. B.}\ }\textbf {\bibinfo {volume} {93}},\ \bibinfo {pages} {169}
  (\bibinfo {year} {2020})}\BibitemShut {NoStop}%
\bibitem [{\citenamefont {Illes}\ \emph {et~al.}(2015)\citenamefont {Illes},
  \citenamefont {Carbotte},\ and\ \citenamefont {Nicol}}]{ICN15}%
  \BibitemOpen
  \bibfield  {author} {\bibinfo {author} {\bibfnamefont {E.}~\bibnamefont
  {Illes}}, \bibinfo {author} {\bibfnamefont {J.~P.}\ \bibnamefont
  {Carbotte}},\ and\ \bibinfo {author} {\bibfnamefont {E.~J.}\ \bibnamefont
  {Nicol}},\ }\href {https://doi.org/10.1103/PhysRevB.92.245410} {\bibfield
  {journal} {\bibinfo  {journal} {Phys. Rev. B}\ }\textbf {\bibinfo {volume}
  {92}},\ \bibinfo {pages} {245410} (\bibinfo {year} {2015})}\BibitemShut
  {NoStop}%
\bibitem [{\citenamefont {Biswas}\ and\ \citenamefont {Ghosh}(2016)}]{BG16}%
  \BibitemOpen
  \bibfield  {author} {\bibinfo {author} {\bibfnamefont {T.}~\bibnamefont
  {Biswas}}\ and\ \bibinfo {author} {\bibfnamefont {T.~K.}\ \bibnamefont
  {Ghosh}},\ }\href {https://doi.org/10.1088/0953-8984/28/49/495302} {\bibfield
   {journal} {\bibinfo  {journal} {Journal of Physics: Condensed Matter}\
  }\textbf {\bibinfo {volume} {28}},\ \bibinfo {pages} {495302} (\bibinfo
  {year} {2016})}\BibitemShut {NoStop}%
\bibitem [{\citenamefont {Illes}\ and\ \citenamefont {Nicol}(2017)}]{IN17}%
  \BibitemOpen
  \bibfield  {author} {\bibinfo {author} {\bibfnamefont {E.}~\bibnamefont
  {Illes}}\ and\ \bibinfo {author} {\bibfnamefont {E.~J.}\ \bibnamefont
  {Nicol}},\ }\href@noop {} {\bibfield  {journal} {\bibinfo  {journal} {Phys.
  Rev. B}\ }\textbf {\bibinfo {volume} {95}},\ \bibinfo {pages} {235432}
  (\bibinfo {year} {2017})}\BibitemShut {NoStop}%
\bibitem [{\citenamefont {Filusch}\ \emph {et~al.}(2020)\citenamefont
  {Filusch}, \citenamefont {Wurl},\ and\ \citenamefont {Fehske}}]{FWF20}%
  \BibitemOpen
  \bibfield  {author} {\bibinfo {author} {\bibfnamefont {A.}~\bibnamefont
  {Filusch}}, \bibinfo {author} {\bibfnamefont {C.}~\bibnamefont {Wurl}},\ and\
  \bibinfo {author} {\bibfnamefont {H.}~\bibnamefont {Fehske}},\ }\href@noop {}
  {\bibfield  {journal} {\bibinfo  {journal} {Eur. Phys. J. B.}\ }\textbf
  {\bibinfo {volume} {90}},\ \bibinfo {pages} {53} (\bibinfo {year}
  {2020})}\BibitemShut {NoStop}%
\bibitem [{\citenamefont {Weekes}\ \emph {et~al.}(2021)\citenamefont {Weekes},
  \citenamefont {Iurov}, \citenamefont {Zhemchuzhna}, \citenamefont {Gumbs},\
  and\ \citenamefont {Huang}}]{WIZGH21}%
  \BibitemOpen
  \bibfield  {author} {\bibinfo {author} {\bibfnamefont {N.}~\bibnamefont
  {Weekes}}, \bibinfo {author} {\bibfnamefont {A.}~\bibnamefont {Iurov}},
  \bibinfo {author} {\bibfnamefont {L.}~\bibnamefont {Zhemchuzhna}}, \bibinfo
  {author} {\bibfnamefont {G.}~\bibnamefont {Gumbs}},\ and\ \bibinfo {author}
  {\bibfnamefont {D.}~\bibnamefont {Huang}},\ }\href
  {https://doi.org/10.1103/PhysRevB.103.165429} {\bibfield  {journal} {\bibinfo
   {journal} {Phys. Rev. B}\ }\textbf {\bibinfo {volume} {103}},\ \bibinfo
  {pages} {165429} (\bibinfo {year} {2021})}\BibitemShut {NoStop}%
\bibitem [{\citenamefont {Cunha}\ \emph {et~al.}(2022)\citenamefont {Cunha},
  \citenamefont {da~Costa}, \citenamefont {Pereira}, \citenamefont {Filho},
  \citenamefont {Van~Duppen},\ and\ \citenamefont {Peeters}}]{CCPFDP22}%
  \BibitemOpen
  \bibfield  {author} {\bibinfo {author} {\bibfnamefont {S.~M.}\ \bibnamefont
  {Cunha}}, \bibinfo {author} {\bibfnamefont {D.~R.}\ \bibnamefont {da~Costa}},
  \bibinfo {author} {\bibfnamefont {J.~M.}\ \bibnamefont {Pereira}}, \bibinfo
  {author} {\bibfnamefont {R.~N.~C.}\ \bibnamefont {Filho}}, \bibinfo {author}
  {\bibfnamefont {B.}~\bibnamefont {Van~Duppen}},\ and\ \bibinfo {author}
  {\bibfnamefont {F.~M.}\ \bibnamefont {Peeters}},\ }\href
  {https://doi.org/10.1103/PhysRevB.105.165402} {\bibfield  {journal} {\bibinfo
   {journal} {Phys. Rev. B}\ }\textbf {\bibinfo {volume} {105}},\ \bibinfo
  {pages} {165402} (\bibinfo {year} {2022})}\BibitemShut {NoStop}%
\bibitem [{\citenamefont {\mbox{Firoz Islam}}\ and\ \citenamefont
  {Datta}(2017)}]{FD17}%
  \BibitemOpen
  \bibfield  {author} {\bibinfo {author} {\bibfnamefont {S.~K.}\ \bibnamefont
  {\mbox{Firoz Islam}}}\ and\ \bibinfo {author} {\bibfnamefont
  {S.}~\bibnamefont {Datta}},\ }\href@noop {} {\bibfield  {journal} {\bibinfo
  {journal} {Phys. Rev. B}\ }\textbf {\bibinfo {volume} {96}},\ \bibinfo
  {pages} {045418} (\bibinfo {year} {2017})}\BibitemShut {NoStop}%
\bibitem [{\citenamefont {Vigh}\ \emph {et~al.}(2013)\citenamefont {Vigh},
  \citenamefont {Oroszl\'any}, \citenamefont {Vajna}, \citenamefont {San-Jose},
  \citenamefont {D\'avid}, \citenamefont {Cserti},\ and\ \citenamefont
  {D\'ora}}]{VOVSDCD13}%
  \BibitemOpen
  \bibfield  {author} {\bibinfo {author} {\bibfnamefont {M.}~\bibnamefont
  {Vigh}}, \bibinfo {author} {\bibfnamefont {L.}~\bibnamefont {Oroszl\'any}},
  \bibinfo {author} {\bibfnamefont {S.}~\bibnamefont {Vajna}}, \bibinfo
  {author} {\bibfnamefont {P.}~\bibnamefont {San-Jose}}, \bibinfo {author}
  {\bibfnamefont {G.}~\bibnamefont {D\'avid}}, \bibinfo {author} {\bibfnamefont
  {J.}~\bibnamefont {Cserti}},\ and\ \bibinfo {author} {\bibfnamefont
  {B.}~\bibnamefont {D\'ora}},\ }\href@noop {} {\bibfield  {journal} {\bibinfo
  {journal} {Phys. Rev. B}\ }\textbf {\bibinfo {volume} {88}},\ \bibinfo
  {pages} {161413} (\bibinfo {year} {2013})}\BibitemShut {NoStop}%
\bibitem [{\citenamefont {Chen}\ \emph {et~al.}(2019)\citenamefont {Chen},
  \citenamefont {Xu}, \citenamefont {Wang}, \citenamefont {Liu},\ and\
  \citenamefont {Ma}}]{CXWLM19}%
  \BibitemOpen
  \bibfield  {author} {\bibinfo {author} {\bibfnamefont {Y.-R.}\ \bibnamefont
  {Chen}}, \bibinfo {author} {\bibfnamefont {Y.}~\bibnamefont {Xu}}, \bibinfo
  {author} {\bibfnamefont {J.}~\bibnamefont {Wang}}, \bibinfo {author}
  {\bibfnamefont {J.-F.}\ \bibnamefont {Liu}},\ and\ \bibinfo {author}
  {\bibfnamefont {Z.}~\bibnamefont {Ma}},\ }\href
  {https://doi.org/10.1103/PhysRevB.99.045420} {\bibfield  {journal} {\bibinfo
  {journal} {Phys. Rev. B}\ }\textbf {\bibinfo {volume} {99}},\ \bibinfo
  {pages} {045420} (\bibinfo {year} {2019})}\BibitemShut {NoStop}%
\bibitem [{\citenamefont {Wang}\ \emph {et~al.}(2017)\citenamefont {Wang},
  \citenamefont {Xu}, \citenamefont {Huang},\ and\ \citenamefont
  {Lai}}]{WXHL17}%
  \BibitemOpen
  \bibfield  {author} {\bibinfo {author} {\bibfnamefont {C.-Z.}\ \bibnamefont
  {Wang}}, \bibinfo {author} {\bibfnamefont {H.-Y.}\ \bibnamefont {Xu}},
  \bibinfo {author} {\bibfnamefont {L.}~\bibnamefont {Huang}},\ and\ \bibinfo
  {author} {\bibfnamefont {Y.-C.}\ \bibnamefont {Lai}},\ }\href@noop {}
  {\bibfield  {journal} {\bibinfo  {journal} {Phys. Rev. B}\ }\textbf {\bibinfo
  {volume} {96}},\ \bibinfo {pages} {115440} (\bibinfo {year}
  {2017})}\BibitemShut {NoStop}%
\bibitem [{\citenamefont {Soni}\ \emph {et~al.}(2020)\citenamefont {Soni},
  \citenamefont {Kaushal}, \citenamefont {Okamoto},\ and\ \citenamefont
  {Dagotto}}]{SKOD20}%
  \BibitemOpen
  \bibfield  {author} {\bibinfo {author} {\bibfnamefont {R.}~\bibnamefont
  {Soni}}, \bibinfo {author} {\bibfnamefont {N.}~\bibnamefont {Kaushal}},
  \bibinfo {author} {\bibfnamefont {S.}~\bibnamefont {Okamoto}},\ and\ \bibinfo
  {author} {\bibfnamefont {E.}~\bibnamefont {Dagotto}},\ }\href
  {https://doi.org/10.1103/PhysRevB.102.045105} {\bibfield  {journal} {\bibinfo
   {journal} {Phys. Rev. B}\ }\textbf {\bibinfo {volume} {102}},\ \bibinfo
  {pages} {045105} (\bibinfo {year} {2020})}\BibitemShut {NoStop}%
\bibitem [{\citenamefont {Wang}\ and\ \citenamefont {Lui}(2011)}]{WL11}%
  \BibitemOpen
  \bibfield  {author} {\bibinfo {author} {\bibfnamefont {Z.~F.}\ \bibnamefont
  {Wang}}\ and\ \bibinfo {author} {\bibfnamefont {F.}~\bibnamefont {Lui}},\
  }\href@noop {} {\bibfield  {journal} {\bibinfo  {journal} {Nanoscale}\
  }\textbf {\bibinfo {volume} {3}},\ \bibinfo {pages} {4201} (\bibinfo {year}
  {2011})}\BibitemShut {NoStop}%
\bibitem [{\citenamefont {Neupert}\ \emph {et~al.}(2011)\citenamefont
  {Neupert}, \citenamefont {Santos}, \citenamefont {Chamon},\ and\
  \citenamefont {Mudry}}]{NSCM11}%
  \BibitemOpen
  \bibfield  {author} {\bibinfo {author} {\bibfnamefont {T.}~\bibnamefont
  {Neupert}}, \bibinfo {author} {\bibfnamefont {L.}~\bibnamefont {Santos}},
  \bibinfo {author} {\bibfnamefont {C.}~\bibnamefont {Chamon}},\ and\ \bibinfo
  {author} {\bibfnamefont {C.}~\bibnamefont {Mudry}},\ }\href
  {https://doi.org/10.1103/PhysRevLett.106.236804} {\bibfield  {journal}
  {\bibinfo  {journal} {Phys. Rev. Lett.}\ }\textbf {\bibinfo {volume} {106}},\
  \bibinfo {pages} {236804} (\bibinfo {year} {2011})}\BibitemShut {NoStop}%
\bibitem [{\citenamefont {Cheng}\ \emph {et~al.}(2021)\citenamefont {Cheng},
  \citenamefont {Zhou}, \citenamefont {Zhou},\ and\ \citenamefont
  {Zhou}}]{CZZZ21}%
  \BibitemOpen
  \bibfield  {author} {\bibinfo {author} {\bibfnamefont {X.}~\bibnamefont
  {Cheng}}, \bibinfo {author} {\bibfnamefont {B.}~\bibnamefont {Zhou}},
  \bibinfo {author} {\bibfnamefont {B.}~\bibnamefont {Zhou}},\ and\ \bibinfo
  {author} {\bibfnamefont {G.}~\bibnamefont {Zhou}},\ }\href
  {https://doi.org/10.1088/1361-648X/abe608} {\bibfield  {journal} {\bibinfo
  {journal} {Journal of Physics: Condensed Matter}\ }\textbf {\bibinfo {volume}
  {33}},\ \bibinfo {pages} {215301} (\bibinfo {year} {2021})}\BibitemShut
  {NoStop}%
\bibitem [{\citenamefont {Dey}\ and\ \citenamefont {Ghosh}(2018)}]{DG18}%
  \BibitemOpen
  \bibfield  {author} {\bibinfo {author} {\bibfnamefont {B.}~\bibnamefont
  {Dey}}\ and\ \bibinfo {author} {\bibfnamefont {T.~K.}\ \bibnamefont
  {Ghosh}},\ }\href {https://doi.org/10.1103/PhysRevB.98.075422} {\bibfield
  {journal} {\bibinfo  {journal} {Phys. Rev. B}\ }\textbf {\bibinfo {volume}
  {98}},\ \bibinfo {pages} {075422} (\bibinfo {year} {2018})}\BibitemShut
  {NoStop}%
\bibitem [{\citenamefont {Mojarro}\ \emph {et~al.}(2020)\citenamefont
  {Mojarro}, \citenamefont {Ibarra-Sierra}, \citenamefont {Sandoval-Santana},
  \citenamefont {Carrillo-Bastos},\ and\ \citenamefont {Naumis}}]{MISCN20}%
  \BibitemOpen
  \bibfield  {author} {\bibinfo {author} {\bibfnamefont {M.~A.}\ \bibnamefont
  {Mojarro}}, \bibinfo {author} {\bibfnamefont {V.~G.}\ \bibnamefont
  {Ibarra-Sierra}}, \bibinfo {author} {\bibfnamefont {J.~C.}\ \bibnamefont
  {Sandoval-Santana}}, \bibinfo {author} {\bibfnamefont {R.}~\bibnamefont
  {Carrillo-Bastos}},\ and\ \bibinfo {author} {\bibfnamefont {G.~G.}\
  \bibnamefont {Naumis}},\ }\href {https://doi.org/10.1103/PhysRevB.101.165305}
  {\bibfield  {journal} {\bibinfo  {journal} {Phys. Rev. B}\ }\textbf {\bibinfo
  {volume} {101}},\ \bibinfo {pages} {165305} (\bibinfo {year}
  {2020})}\BibitemShut {NoStop}%
\bibitem [{\citenamefont {Iurov}\ \emph {et~al.}(2022)\citenamefont {Iurov},
  \citenamefont {Zhemchuzhna}, \citenamefont {Gumbs}, \citenamefont {Huang},\
  and\ \citenamefont {Fekete}}]{IZGHF22}%
  \BibitemOpen
  \bibfield  {author} {\bibinfo {author} {\bibfnamefont {A.}~\bibnamefont
  {Iurov}}, \bibinfo {author} {\bibfnamefont {L.}~\bibnamefont {Zhemchuzhna}},
  \bibinfo {author} {\bibfnamefont {G.}~\bibnamefont {Gumbs}}, \bibinfo
  {author} {\bibfnamefont {D.}~\bibnamefont {Huang}},\ and\ \bibinfo {author}
  {\bibfnamefont {P.}~\bibnamefont {Fekete}},\ }\href
  {https://doi.org/10.1103/PhysRevB.105.115309} {\bibfield  {journal} {\bibinfo
   {journal} {Phys. Rev. B}\ }\textbf {\bibinfo {volume} {105}},\ \bibinfo
  {pages} {115309} (\bibinfo {year} {2022})}\BibitemShut {NoStop}%
\bibitem [{\citenamefont {Cheng}\ and\ \citenamefont {Xianlong}(2022)}]{CX22}%
  \BibitemOpen
  \bibfield  {author} {\bibinfo {author} {\bibfnamefont {S.}~\bibnamefont
  {Cheng}}\ and\ \bibinfo {author} {\bibfnamefont {G.}~\bibnamefont
  {Xianlong}},\ }\href {https://doi.org/10.1103/PhysRevResearch.4.033194}
  {\bibfield  {journal} {\bibinfo  {journal} {Phys. Rev. Research}\ }\textbf
  {\bibinfo {volume} {4}},\ \bibinfo {pages} {033194} (\bibinfo {year}
  {2022})}\BibitemShut {NoStop}%
\bibitem [{\citenamefont {Dey}\ and\ \citenamefont {Ghosh}(2019)}]{DG19}%
  \BibitemOpen
  \bibfield  {author} {\bibinfo {author} {\bibfnamefont {B.}~\bibnamefont
  {Dey}}\ and\ \bibinfo {author} {\bibfnamefont {T.~K.}\ \bibnamefont
  {Ghosh}},\ }\href {https://doi.org/10.1103/PhysRevB.99.205429} {\bibfield
  {journal} {\bibinfo  {journal} {Phys. Rev. B}\ }\textbf {\bibinfo {volume}
  {99}},\ \bibinfo {pages} {205429} (\bibinfo {year} {2019})}\BibitemShut
  {NoStop}%
\bibitem [{\citenamefont {Xu}\ \emph {et~al.}(2017)\citenamefont {Xu},
  \citenamefont {Huang}, \citenamefont {Huang},\ and\ \citenamefont
  {Lai}}]{XHHL17}%
  \BibitemOpen
  \bibfield  {author} {\bibinfo {author} {\bibfnamefont {H.-Y.}\ \bibnamefont
  {Xu}}, \bibinfo {author} {\bibfnamefont {L.}~\bibnamefont {Huang}}, \bibinfo
  {author} {\bibfnamefont {D.}~\bibnamefont {Huang}},\ and\ \bibinfo {author}
  {\bibfnamefont {Y.-C.}\ \bibnamefont {Lai}},\ }\href@noop {} {\bibfield
  {journal} {\bibinfo  {journal} {Phys. Rev. B}\ }\textbf {\bibinfo {volume}
  {96}},\ \bibinfo {pages} {045412} (\bibinfo {year} {2017})}\BibitemShut
  {NoStop}%
\bibitem [{\citenamefont {Bouhadida}\ \emph {et~al.}(2020)\citenamefont
  {Bouhadida}, \citenamefont {Mandhour},\ and\ \citenamefont
  {Charfi-Kaddour}}]{BMCK20}%
  \BibitemOpen
  \bibfield  {author} {\bibinfo {author} {\bibfnamefont {F.}~\bibnamefont
  {Bouhadida}}, \bibinfo {author} {\bibfnamefont {L.}~\bibnamefont
  {Mandhour}},\ and\ \bibinfo {author} {\bibfnamefont {S.}~\bibnamefont
  {Charfi-Kaddour}},\ }\href {https://doi.org/10.1103/PhysRevB.102.075443}
  {\bibfield  {journal} {\bibinfo  {journal} {Phys. Rev. B}\ }\textbf {\bibinfo
  {volume} {102}},\ \bibinfo {pages} {075443} (\bibinfo {year}
  {2020})}\BibitemShut {NoStop}%
\bibitem [{\citenamefont {Zeng}\ and\ \citenamefont {Shen}(2022)}]{ZS22}%
  \BibitemOpen
  \bibfield  {author} {\bibinfo {author} {\bibfnamefont {W.}~\bibnamefont
  {Zeng}}\ and\ \bibinfo {author} {\bibfnamefont {R.}~\bibnamefont {Shen}},\
  }\href {https://doi.org/10.1103/PhysRevB.106.094503} {\bibfield  {journal}
  {\bibinfo  {journal} {Phys. Rev. B}\ }\textbf {\bibinfo {volume} {106}},\
  \bibinfo {pages} {094503} (\bibinfo {year} {2022})}\BibitemShut {NoStop}%
\bibitem [{\citenamefont {Filusch}\ \emph {et~al.}(2021)\citenamefont
  {Filusch}, \citenamefont {Bishop}, \citenamefont {Saxena}, \citenamefont
  {Wellein},\ and\ \citenamefont {Fehske}}]{FBSWH21}%
  \BibitemOpen
  \bibfield  {author} {\bibinfo {author} {\bibfnamefont {A.}~\bibnamefont
  {Filusch}}, \bibinfo {author} {\bibfnamefont {A.~R.}\ \bibnamefont {Bishop}},
  \bibinfo {author} {\bibfnamefont {A.}~\bibnamefont {Saxena}}, \bibinfo
  {author} {\bibfnamefont {G.}~\bibnamefont {Wellein}},\ and\ \bibinfo {author}
  {\bibfnamefont {H.}~\bibnamefont {Fehske}},\ }\href
  {https://doi.org/10.1103/PhysRevB.103.165114} {\bibfield  {journal} {\bibinfo
   {journal} {Phys. Rev. B}\ }\textbf {\bibinfo {volume} {103}},\ \bibinfo
  {pages} {165114} (\bibinfo {year} {2021})}\BibitemShut {NoStop}%
\bibitem [{\citenamefont {Sun}\ \emph {et~al.}(2022)\citenamefont {Sun},
  \citenamefont {Liu}, \citenamefont {Du},\ and\ \citenamefont {Guo}}]{SLDG22}%
  \BibitemOpen
  \bibfield  {author} {\bibinfo {author} {\bibfnamefont {J.}~\bibnamefont
  {Sun}}, \bibinfo {author} {\bibfnamefont {T.}~\bibnamefont {Liu}}, \bibinfo
  {author} {\bibfnamefont {Y.}~\bibnamefont {Du}},\ and\ \bibinfo {author}
  {\bibfnamefont {H.}~\bibnamefont {Guo}},\ }\href
  {https://doi.org/10.1103/PhysRevB.106.155417} {\bibfield  {journal} {\bibinfo
   {journal} {Phys. Rev. B}\ }\textbf {\bibinfo {volume} {106}},\ \bibinfo
  {pages} {155417} (\bibinfo {year} {2022})}\BibitemShut {NoStop}%
\bibitem [{\citenamefont {Groth}\ \emph {et~al.}(2014)\citenamefont {Groth},
  \citenamefont {Wimmer}, \citenamefont {Akhmerov},\ and\ \citenamefont
  {Waintal}}]{GWAW14}%
  \BibitemOpen
  \bibfield  {author} {\bibinfo {author} {\bibfnamefont {C.~W.}\ \bibnamefont
  {Groth}}, \bibinfo {author} {\bibfnamefont {M.}~\bibnamefont {Wimmer}},
  \bibinfo {author} {\bibfnamefont {A.~R.}\ \bibnamefont {Akhmerov}},\ and\
  \bibinfo {author} {\bibfnamefont {X.}~\bibnamefont {Waintal}},\ }\href@noop
  {} {\bibfield  {journal} {\bibinfo  {journal} {New Journal of Physics}\
  }\textbf {\bibinfo {volume} {16}},\ \bibinfo {pages} {063065} (\bibinfo
  {year} {2014})}\BibitemShut {NoStop}%
\bibitem [{\citenamefont {Qi}\ \emph {et~al.}(2017)\citenamefont {Qi},
  \citenamefont {Shi}, \citenamefont {Liu},\ and\ \citenamefont
  {Kruchinin}}]{QSLK17}%
  \BibitemOpen
  \bibfield  {author} {\bibinfo {author} {\bibfnamefont {F.}~\bibnamefont
  {Qi}}, \bibinfo {author} {\bibfnamefont {X.-T.}\ \bibnamefont {Shi}},
  \bibinfo {author} {\bibfnamefont {F.-F.}\ \bibnamefont {Liu}},\ and\ \bibinfo
  {author} {\bibfnamefont {D.~V.}\ \bibnamefont {Kruchinin}},\ }\href@noop {}
  {\bibfield  {journal} {\bibinfo  {journal} {Journal of Applied Analysis and
  Computation}\ }\textbf {\bibinfo {volume} {7}},\ \bibinfo {pages} {857}
  (\bibinfo {year} {2017})}\BibitemShut {NoStop}%
\bibitem [{\citenamefont {Zwillinger}\ \emph {et~al.}(2014)\citenamefont
  {Zwillinger}, \citenamefont {Moll}, \citenamefont {Gradshteyn},\ and\
  \citenamefont {Ryzhik}}]{ZMGR14}%
  \BibitemOpen
  \bibinfo {editor} {\bibfnamefont {D.}~\bibnamefont {Zwillinger}}, \bibinfo
  {editor} {\bibfnamefont {V.}~\bibnamefont {Moll}}, \bibinfo {editor}
  {\bibfnamefont {I.}~\bibnamefont {Gradshteyn}},\ and\ \bibinfo {editor}
  {\bibfnamefont {I.}~\bibnamefont {Ryzhik}},\ eds.,\ \href@noop {} {\emph
  {\bibinfo {title} {Table of Integrals, Series, and Products}}},\ \bibinfo
  {edition} {eighth edition}\ ed.\ (\bibinfo  {publisher} {Academic Press},\
  \bibinfo {address} {Boston},\ \bibinfo {year} {2014})\BibitemShut {NoStop}%
\bibitem [{\citenamefont {Pereira}\ and\ \citenamefont
  {Castro~Neto}(2009)}]{PN09}%
  \BibitemOpen
  \bibfield  {author} {\bibinfo {author} {\bibfnamefont {V.~M.}\ \bibnamefont
  {Pereira}}\ and\ \bibinfo {author} {\bibfnamefont {A.~H.}\ \bibnamefont
  {Castro~Neto}},\ }\href {https://doi.org/10.1103/PhysRevLett.103.046801}
  {\bibfield  {journal} {\bibinfo  {journal} {Phys. Rev. Lett.}\ }\textbf
  {\bibinfo {volume} {103}},\ \bibinfo {pages} {046801} (\bibinfo {year}
  {2009})}\BibitemShut {NoStop}%
\bibitem [{\citenamefont {Sloan}\ \emph {et~al.}(2013)\citenamefont {Sloan},
  \citenamefont {Sanjuan}, \citenamefont {Wang}, \citenamefont {Horvath},\ and\
  \citenamefont {Barraza-Lopez}}]{SSWHB13}%
  \BibitemOpen
  \bibfield  {author} {\bibinfo {author} {\bibfnamefont {J.~V.}\ \bibnamefont
  {Sloan}}, \bibinfo {author} {\bibfnamefont {A.~A.~P.}\ \bibnamefont
  {Sanjuan}}, \bibinfo {author} {\bibfnamefont {Z.}~\bibnamefont {Wang}},
  \bibinfo {author} {\bibfnamefont {C.}~\bibnamefont {Horvath}},\ and\ \bibinfo
  {author} {\bibfnamefont {S.}~\bibnamefont {Barraza-Lopez}},\ }\href
  {https://doi.org/10.1103/PhysRevB.87.155436} {\bibfield  {journal} {\bibinfo
  {journal} {Phys. Rev. B}\ }\textbf {\bibinfo {volume} {87}},\ \bibinfo
  {pages} {155436} (\bibinfo {year} {2013})}\BibitemShut {NoStop}%
\bibitem [{\citenamefont {Kohler}\ \emph {et~al.}(2005)\citenamefont {Kohler},
  \citenamefont {Lehmann},\ and\ \citenamefont {H{\"a}nggi}}]{KLH05}%
  \BibitemOpen
  \bibfield  {author} {\bibinfo {author} {\bibfnamefont {S.}~\bibnamefont
  {Kohler}}, \bibinfo {author} {\bibfnamefont {J.}~\bibnamefont {Lehmann}},\
  and\ \bibinfo {author} {\bibfnamefont {P.}~\bibnamefont {H{\"a}nggi}},\
  }\href {https://doi.org/https://doi.org/10.1016/j.physrep.2004.11.002}
  {\bibfield  {journal} {\bibinfo  {journal} {Physics Reports}\ }\textbf
  {\bibinfo {volume} {406}},\ \bibinfo {pages} {379} (\bibinfo {year}
  {2005})}\BibitemShut {NoStop}%
\bibitem [{\citenamefont {Fruchart}\ \emph {et~al.}(2016)\citenamefont
  {Fruchart}, \citenamefont {Delplace}, \citenamefont {Weston}, \citenamefont
  {Waintal},\ and\ \citenamefont {Carpentier}}]{FDWWC16}%
  \BibitemOpen
  \bibfield  {author} {\bibinfo {author} {\bibfnamefont {M.}~\bibnamefont
  {Fruchart}}, \bibinfo {author} {\bibfnamefont {P.}~\bibnamefont {Delplace}},
  \bibinfo {author} {\bibfnamefont {J.}~\bibnamefont {Weston}}, \bibinfo
  {author} {\bibfnamefont {X.}~\bibnamefont {Waintal}},\ and\ \bibinfo {author}
  {\bibfnamefont {D.}~\bibnamefont {Carpentier}},\ }\href
  {https://doi.org/https://doi.org/10.1016/j.physe.2015.09.035} {\bibfield
  {journal} {\bibinfo  {journal} {Physica E: Low-dimensional Systems and
  Nanostructures}\ }\textbf {\bibinfo {volume} {75}},\ \bibinfo {pages} {287}
  (\bibinfo {year} {2016})}\BibitemShut {NoStop}%
\bibitem [{\citenamefont {Kitagawa}\ \emph {et~al.}(2011)\citenamefont
  {Kitagawa}, \citenamefont {Oka}, \citenamefont {Brataas}, \citenamefont
  {Fu},\ and\ \citenamefont {Demler}}]{KOBFD11}%
  \BibitemOpen
  \bibfield  {author} {\bibinfo {author} {\bibfnamefont {T.}~\bibnamefont
  {Kitagawa}}, \bibinfo {author} {\bibfnamefont {T.}~\bibnamefont {Oka}},
  \bibinfo {author} {\bibfnamefont {A.}~\bibnamefont {Brataas}}, \bibinfo
  {author} {\bibfnamefont {L.}~\bibnamefont {Fu}},\ and\ \bibinfo {author}
  {\bibfnamefont {E.}~\bibnamefont {Demler}},\ }\href
  {https://doi.org/10.1103/PhysRevB.84.235108} {\bibfield  {journal} {\bibinfo
  {journal} {Phys. Rev. B}\ }\textbf {\bibinfo {volume} {84}},\ \bibinfo
  {pages} {235108} (\bibinfo {year} {2011})}\BibitemShut {NoStop}%
\bibitem [{\citenamefont {Yap}\ \emph {et~al.}(2017)\citenamefont {Yap},
  \citenamefont {Zhou}, \citenamefont {Wang},\ and\ \citenamefont
  {Gong}}]{YZWG17}%
  \BibitemOpen
  \bibfield  {author} {\bibinfo {author} {\bibfnamefont {H.~H.}\ \bibnamefont
  {Yap}}, \bibinfo {author} {\bibfnamefont {L.}~\bibnamefont {Zhou}}, \bibinfo
  {author} {\bibfnamefont {J.-S.}\ \bibnamefont {Wang}},\ and\ \bibinfo
  {author} {\bibfnamefont {J.}~\bibnamefont {Gong}},\ }\href
  {https://doi.org/10.1103/PhysRevB.96.165443} {\bibfield  {journal} {\bibinfo
  {journal} {Phys. Rev. B}\ }\textbf {\bibinfo {volume} {96}},\ \bibinfo
  {pages} {165443} (\bibinfo {year} {2017})}\BibitemShut {NoStop}%
\bibitem [{\citenamefont {Lewenkopf}\ and\ \citenamefont
  {Mucciolo}(2013)}]{LM13}%
  \BibitemOpen
  \bibfield  {author} {\bibinfo {author} {\bibfnamefont {C.~H.}\ \bibnamefont
  {Lewenkopf}}\ and\ \bibinfo {author} {\bibfnamefont {E.~R.}\ \bibnamefont
  {Mucciolo}},\ }\href {https://doi.org/10.1007/s10825-013-0458-7} {\bibfield
  {journal} {\bibinfo  {journal} {Journal of Computational Electronics}\
  }\textbf {\bibinfo {volume} {12}},\ \bibinfo {pages} {203} (\bibinfo {year}
  {2013})}\BibitemShut {NoStop}%
\bibitem [{\citenamefont {Thorgilsson}\ \emph {et~al.}(2014)\citenamefont
  {Thorgilsson}, \citenamefont {Viktorsson},\ and\ \citenamefont
  {Erlingsson}}]{TVE14}%
  \BibitemOpen
  \bibfield  {author} {\bibinfo {author} {\bibfnamefont {G.}~\bibnamefont
  {Thorgilsson}}, \bibinfo {author} {\bibfnamefont {G.}~\bibnamefont
  {Viktorsson}},\ and\ \bibinfo {author} {\bibfnamefont {S.}~\bibnamefont
  {Erlingsson}},\ }\href
  {https://doi.org/https://doi.org/10.1016/j.jcp.2013.12.054} {\bibfield
  {journal} {\bibinfo  {journal} {Journal of Computational Physics}\ }\textbf
  {\bibinfo {volume} {261}},\ \bibinfo {pages} {256} (\bibinfo {year}
  {2014})}\BibitemShut {NoStop}%
\bibitem [{\citenamefont {Stegmann}\ and\ \citenamefont {Szpak}(2018)}]{SS19}%
  \BibitemOpen
  \bibfield  {author} {\bibinfo {author} {\bibfnamefont {T.}~\bibnamefont
  {Stegmann}}\ and\ \bibinfo {author} {\bibfnamefont {N.}~\bibnamefont
  {Szpak}},\ }\href {https://doi.org/10.1088/2053-1583/aaea8d} {\bibfield
  {journal} {\bibinfo  {journal} {2D Materials}\ }\textbf {\bibinfo {volume}
  {6}},\ \bibinfo {pages} {015024} (\bibinfo {year} {2018})}\BibitemShut
  {NoStop}%
  \bibitem [{\citenamefont {B\"auml}\ \emph {et~al.}(1998)\citenamefont
  {B\"auml}, \citenamefont {Wellein},\ and\ \citenamefont {Fehske}}]{BWF98}%
  \BibitemOpen
  \bibfield  {author} {\bibinfo {author} {\bibfnamefont {B.}~\bibnamefont
  {B\"auml}}, \bibinfo {author} {\bibfnamefont {G.}~\bibnamefont {Wellein}},\
  and\ \bibinfo {author} {\bibfnamefont {H.}~\bibnamefont {Fehske}},\ }\href
  {https://doi.org/10.1103/PhysRevB.58.3663} {\bibfield  {journal} {\bibinfo
  {journal} {Phys. Rev. B}\ }\textbf {\bibinfo {volume} {58}},\ \bibinfo
  {pages} {3663} (\bibinfo {year} {1998})}\BibitemShut {NoStop}%
\bibitem [{\citenamefont {Wei{\ss}e}\ \emph {et~al.}(2006)\citenamefont
  {Wei{\ss}e}, \citenamefont {Wellein}, \citenamefont {Alvermann},\ and\
  \citenamefont {Fehske}}]{WWAF06}%
  \BibitemOpen
  \bibfield  {author} {\bibinfo {author} {\bibfnamefont {A.}~\bibnamefont
  {Wei{\ss}e}}, \bibinfo {author} {\bibfnamefont {G.}~\bibnamefont {Wellein}},
  \bibinfo {author} {\bibfnamefont {A.}~\bibnamefont {Alvermann}},\ and\
  \bibinfo {author} {\bibfnamefont {H.}~\bibnamefont {Fehske}},\ }\href@noop {}
  {\bibfield  {journal} {\bibinfo  {journal} {Rev. Mod. Phys.}\ }\textbf
  {\bibinfo {volume} {78}},\ \bibinfo {pages} {275} (\bibinfo {year}
  {2006})}\BibitemShut {NoStop}%
\bibitem [{\citenamefont {Haug}\ and\ \citenamefont {Jauho}(2008)}]{HJ08}%
  \BibitemOpen
  \bibfield  {author} {\bibinfo {author} {\bibfnamefont {H.}~\bibnamefont
  {Haug}}\ and\ \bibinfo {author} {\bibfnamefont {A.-P.}\ \bibnamefont
  {Jauho}},\ }\href@noop {} {\emph {\bibinfo {title} {Quantum Kinetics in
  Transport and Optics of Semiconductors}}}\ (\bibinfo  {publisher}
  {Springer},\ \bibinfo {address} {Berlin Heidelberg New-York},\ \bibinfo
  {year} {2008})\BibitemShut {NoStop}%
\bibitem [{\citenamefont {{Wang, J.-S.}}\ \emph {et~al.}(2008)\citenamefont
  {{Wang, J.-S.}}, \citenamefont {{Wang, J.}},\ and\ \citenamefont {{L\"u, J.
  T.}}}]{WWL08}%
  \BibitemOpen
  \bibfield  {author} {\bibinfo {author} {\bibnamefont {{Wang, J.-S.}}},
  \bibinfo {author} {\bibnamefont {{Wang, J.}}},\ and\ \bibinfo {author}
  {\bibnamefont {{L\"u, J. T.}}},\ }\href
  {https://doi.org/10.1140/epjb/e2008-00195-8} {\bibfield  {journal} {\bibinfo
  {journal} {Eur. Phys. J. B}\ }\textbf {\bibinfo {volume} {62}},\ \bibinfo
  {pages} {381} (\bibinfo {year} {2008})}\BibitemShut {NoStop}%
\bibitem [{\citenamefont {Magnus}(1954)}]{M54}%
  \BibitemOpen
  \bibfield  {author} {\bibinfo {author} {\bibfnamefont {W.}~\bibnamefont
  {Magnus}},\ }\href@noop {} {\bibfield  {journal} {\bibinfo  {journal}
  {Communications on Pure and Applied Mathematics}\ }\textbf {\bibinfo {volume}
  {7}},\ \bibinfo {pages} {649} (\bibinfo {year} {1954})}\BibitemShut {NoStop}%
\bibitem [{\citenamefont {Jamotte}\ \emph {et~al.}(2022)\citenamefont
  {Jamotte}, \citenamefont {Goldman},\ and\ \citenamefont
  {Di~Liberto}}]{JGL22}%
  \BibitemOpen
  \bibfield  {author} {\bibinfo {author} {\bibfnamefont {M.}~\bibnamefont
  {Jamotte}}, \bibinfo {author} {\bibfnamefont {N.}~\bibnamefont {Goldman}},\
  and\ \bibinfo {author} {\bibfnamefont {M.}~\bibnamefont {Di~Liberto}},\
  }\href {https://doi.org/10.1038/s42005-022-00802-9} {\bibfield  {journal}
  {\bibinfo  {journal} {Commun Phys.}\ }\textbf {\bibinfo {volume} {5}},\
  \bibinfo {pages} {30} (\bibinfo {year} {2022})}\BibitemShut {NoStop}%
\bibitem [{\citenamefont {Weinberg}\ \emph {et~al.}(2016)\citenamefont
  {Weinberg}, \citenamefont {Staarmann}, \citenamefont {{\"O}lschl{\"a}ger},
  \citenamefont {Simonet},\ and\ \citenamefont {Sengstock}}]{WSOESS16}%
  \BibitemOpen
  \bibfield  {author} {\bibinfo {author} {\bibfnamefont {M.}~\bibnamefont
  {Weinberg}}, \bibinfo {author} {\bibfnamefont {C.}~\bibnamefont {Staarmann}},
  \bibinfo {author} {\bibfnamefont {C.}~\bibnamefont {{\"O}lschl{\"a}ger}},
  \bibinfo {author} {\bibfnamefont {J.}~\bibnamefont {Simonet}},\ and\ \bibinfo
  {author} {\bibfnamefont {K.}~\bibnamefont {Sengstock}},\ }\href
  {https://doi.org/10.1088/2053-1583/3/2/024005} {\bibfield  {journal}
  {\bibinfo  {journal} {2D Materials}\ }\textbf {\bibinfo {volume} {3}},\
  \bibinfo {pages} {024005} (\bibinfo {year} {2016})}\BibitemShut {NoStop}%
\bibitem [{\citenamefont {Jotzu}\ \emph {et~al.}(2014)\citenamefont {Jotzu},
  \citenamefont {Messer}, \citenamefont {Desbuquois}, \citenamefont {Lebrat},
  \citenamefont {Uehlinger}, \citenamefont {Greif},\ and\ \citenamefont
  {Esslinger}}]{JMDLUGE14}%
  \BibitemOpen
  \bibfield  {author} {\bibinfo {author} {\bibfnamefont {G.}~\bibnamefont
  {Jotzu}}, \bibinfo {author} {\bibfnamefont {M.}~\bibnamefont {Messer}},
  \bibinfo {author} {\bibfnamefont {R.}~\bibnamefont {Desbuquois}}, \bibinfo
  {author} {\bibfnamefont {M.}~\bibnamefont {Lebrat}}, \bibinfo {author}
  {\bibfnamefont {T.}~\bibnamefont {Uehlinger}}, \bibinfo {author}
  {\bibfnamefont {D.}~\bibnamefont {Greif}},\ and\ \bibinfo {author}
  {\bibfnamefont {T.}~\bibnamefont {Esslinger}},\ }\href@noop {} {\bibfield
  {journal} {\bibinfo  {journal} {Nature}\ }\textbf {\bibinfo {volume} {515}},\
  \bibinfo {pages} {237} (\bibinfo {year} {2014})}\BibitemShut {NoStop}%
\end{thebibliography}
\bibliographystyle{apsrev4-2}

\end{document}